\documentclass[a4paper,fleqn]{cas-dc} 

\makeatletter
\g@addto@macro\@floatboxreset{\setlength{\linewidth}{0.5\textwidth}\centering}
\makeatother

\usepackage[numbers]{natbib}
\usepackage{hyperref}
\usepackage{doi}

\usepackage{lineno} 
\usepackage{makecell}
\usepackage{float}
\usepackage{stfloats}
\usepackage{placeins}
\usepackage{graphicx}
\usepackage[T1]{fontenc}
\usepackage{subcaption}
\usepackage{amsmath}
\usepackage{mathtools} 
\usepackage{amssymb}   
\usepackage{siunitx}   
\usepackage{longtable,booktabs,siunitx}
\usepackage{hyperref} 

\DeclareSIUnit{\photons}{photons}
\DeclareSIUnit{\ph}{ph}
\DeclareSIUnit{\MeV}{\mega\electronvolt}
\DeclareSIUnit{\MeValpha}{\MeV\textsubscript{\alpha}} 

\usepackage[acronym, nohypertypes={acronym}]{glossaries}

\graphicspath{{Graphs/crlb simulations/}{Graphs/CRT_FWHM_plots/}{Graphs/Energy Spectra/}{Graphs/Fusion drawings/}{Graphs/Position_Reconstruction/Measurement/}{Graphs/Position_Reconstruction/Simulation/}{Graphs/samples/}{Graphs/Single Photon Light Yield/}{Graphs/tcspc/}{Graphs/}}
        
\makeglossaries
\newacronym{crlb}{CRLB}{Cramèr-Rao lower bound}
\newacronym{tcspc}{TCSPC}{time-correlated single-photon counting}
\newacronym{gan}{GaN:Si}{silicon-doped gallium nitride}
\newacronym{zno}{ZnO:Ga}{gallium-doped zinc oxide}
\newacronym{yap}{YAP:Ce}{cerium-doped yttrium aluminium perovskite}
\newacronym{ctr}{CTR}{coincidence timing resolution}
\newacronym{dtr}{DTR}{detector timing resolution}
\newacronym{cfd}{CFD}{constant fraction discrimination}
\newacronym{tts}{TTS}{transit-time spread}
\newacronym{pmt}{PMT}{photomultiplier tube}
\newacronym{qe}{QE}{quantum efficiency}
\newacronym{api}{API}{Associated Particle Imaging}
\newacronym{p.d.f.}{p.d.f.}{probability density function}
\newacronym{fwhm}{FWHM}{full width at half maximum}
\newacronym{movpe}{MOVPE}{metal-organic vapor-phase epitaxy}
\newacronym{am241}{$^{241}$Am}{Americium 241}
\newacronym{pl}{PL}{photoluminescence}
\newacronym{mcp}{MCP-PMT}{microchannel plate PMT}
\newacronym{irf}{IRF}{instrument response function}
\newacronym{pmma}{PMMA}{polymethyl methacrylate}
\newacronym{pet}{PET}{Positron-emission tomography}
\newacronym{ct}{CT}{Computed Tomography}

\def\tsc#1{\csdef{#1}{\textsc{\lowercase{#1}}\xspace}}
\tsc{WGM}
\tsc{QE}
\tsc{EP}
\tsc{PMS}
\tsc{BEC}
\tsc{DE}

\begin{document}
\let\WriteBookmarks\relax
\def\floatpagepagefraction{1}
\def\textpagefraction{.001}
\shortauthors{J. Meyer et~al.}

\shorttitle{Characterization of GaN:Si and ZnO:Ga}
\title[mode = title]{Characterization of GaN:Si and ZnO:Ga for position-resolved fast timing applications}

\author[1,3]{Julius Meyer}
\cormark[1]

\ead{jmeyer@ikp.tu-darmstadt.de} 

\author[2]{Joshua W. Cates} 

\author[2]{Woon-Seng Choong} 

\author[1,4]{Juan Cristhian Luque Gutierrez}  

\author[2]{Federico Moretti} 

\author[5]{Ryan Pavlovsky} 

\author[1]{Mauricio Ayllon Unzueta}  
\cormark[2]
\ead{mayllonu@lbl.gov} 

\author[2]{Weronika W. Wolszczak} 

\author[3]{Markus Roth} 

\author[1]{Arun Persaud}  

\affiliation[1]{organization={Accelerator Technology \& Applied Physics Division, E.O. Lawrence Berkeley National Laboratory},
                city={Berkeley},
                postcode={CA 94720}, 
                country={USA}}

\affiliation[2]{organization={Nuclear Science Division, E.O. Lawrence Berkeley National Laboratory},
                city={Berkeley},
                postcode={CA 94720}, 
                country={USA}}
                
\affiliation[3]{organization={Institute for Nuclear Physics, Technical University of Darmstadt},
                city={Darmstadt},
                postcode={64289}, 
                country={Germany}}

\affiliation[4]{organization={North Carolina State University},
                city={Raleigh},
                postcode={NC 27695}, 
                country={USA}}

\affiliation[5]{organization={Gamma Reality Inc.},
                city={Richmond},
                postcode={CA 94804}, 
                country={USA}}

\cortext[cor1]{Corresponding authors}

\begin{abstract}
We present the characterization of two fast, crystalline inorganic scintillators, \gls{gan} and \gls{zno}, and compare their performance with \gls{yap} for in-vacuum alpha-detection applications that require high-performance timing, position, and energy resolution, such as 3D elemental mapping, medical imaging, and homeland security applications. In this paper, we propose \gls{zno} and \gls{gan} as high-performance drop-in replacements for the alpha detector in \gls{api} systems. However, the results reported here also have wide applicability. Prior work has reported on polycrystalline forms of \gls{zno}, which suffer from self-absorption or non-spectroscopic alpha energy spectra, and, to our knowledge, \gls{gan} has not been proposed to be used in \gls{api} systems. We present room-temperature scintillation time constants obtained via X-ray-induced \acrlong{tcspc} for both proposed materials. They both exhibit exceedingly fast rise times of \qty{\le 15}{\pico\second}, and high brightness \qty{\ge 1000}{ph/\mega\electronvolt}$_{\alpha}$ with resolved alpha-peaks. Single-crystal \gls{zno} and single-crystal \gls{gan} yield single-component decays of \qty{805}{\pico\second} and \qty{32}{\pico\second} respectively. Using a plastic scintillator reference setup, \gls{ctr} and \gls{dtr} measurements demonstrate a $\ge$$3\times$ improvement in timing resolution compared to traditional \gls{yap}. \gls{gan} and \gls{zno} exhibit \qty{35 \pm 9}{\pico\second} and \qty{49 \pm 5}{\pico\second} \gls{dtr} respectively, compared to \qty{144 \pm 2}{\pico\second} for conventional, single-crystal \gls{yap}. Finally, we evaluate their position resolution in an experimental setup designed for \gls{api} and measure better than \qty{0.2}{\milli\meter} for \gls{yap} and approximately \qty{1}{\milli\meter} for \gls{gan}. We obtain a position resolution of \qty{0.3}{\milli\meter} for \gls{zno} from simulations. We also present alpha-induced ionoluminescence emission spectra that reveal direct, red-shifted near-bandgap emission.
\end{abstract}



\maketitle

\glsresetall

\section{Introduction and Background}

High-performance radiation detection systems, in particular those achieving fast timing, are of interest to many application domains, from 3D elemental mapping in soils or for planetary exploration to medical imaging, and the detection and imaging of landmines. In medical imaging, \gls{pet} and \gls{ct} have aspirational timing-resolution targets of \qty{\le 10}{\pico\second} \gls{ctr} \gls{fwhm} \cite{gundacker2019_SiPM_tof-pet}. Reaching this benchmark requires scintillators that are both bright and fast. In \gls{pet} and \gls{ct}, improvements in scintillator materials for X-ray and gamma-ray detection are key, and their properties, in a well-engineered system, determine the fundamental limits of performance. Enhancing the scintillator performance would enable, for example, high-precision tumor detection with lower radiation dose for the patient. For 3D elemental mapping, techniques such as \gls{api} can be applied. For this technique, fast scintillators with good energy and position resolution for alpha particle detection are needed to provide trajectory information of the associated particles (the neutrons). At the same time, the scintillator needs to have low sensitivity to gamma-rays and X-rays. Strong candidate materials exhibit fast \gls{dtr} performance and short decay constants to enhance \gls{api} image resolution and reduce pile-up, which can contribute to rate improvements of 10$\times$ \cite{ayllon2021_api_3d_recon}. Such improvements would enable high-sensitivity scans for homeland security applications \cite{ji2015_api_SNM}, material composition analyses \cite{ayllon2021_api_3d_recon}, cross-section measurements, planetary exploration missions \cite{inspect3r}, and field-deployed humanitarian de-mining \cite{PavlovskySORMAWest2025}. Although \gls{yap} is widely used in current \gls{api} systems, its intrinsic scintillation properties impose fundamental limits on timing precision, rate capability, and photon statistics, which collectively constrain achievable timing and position resolution. In contrast, \gls{gan} and \gls{zno} are inorganic semiconductor scintillators with near-band-gap emission and sub-nano\-second decay times, making them intrinsically better aligned with the demands of high-rate, high-resolution \gls{api}. In this work, we focus on alpha-particle detection using \gls{gan} and \gls{zno}, and compare their performance with \gls{yap}. We measure (i) rise and decay times, (ii) light yield and emission spectra under alpha particle irradiation, (iii) coincidence timing resolution, and (iv) position resolution capabilities of our samples.
\\

\subsection{Rise and Decay Times}
In previous work, primary decay times were reported to be less than \qty{1}{\nano\second} for \gls{gan} (e.g., \cite{schenk2009cathodoluminescence_gan_si}) and \gls{zno} (e.g., \cite{lehmann1966edge}). While unintentionally doped GaN was studied for its relatively slow yellow luminescence scintillation component, e.g., in \citet{yanagida2021photoluminescence_Gan}, \citet{pittet2009_doping_effect_Si_in_gan} have shown that for silicon concentrations on the order of \qty{1e19}{\per\cubic\centi\meter} the ultra-fast direct band-gap emission becomes dominant. For \gls{gan} at \qty{1e19}{\per\cubic\centi\meter}, the decay constant of the fast \qty{365}{\nano\meter} scintillation emission was reported to be \qty{500}{\pico\second} by \citet{schenk2008gan_si_doping}. For \gls{zno}, \citet{cates2013zno_ga} measured the decay time of their polycrystalline \gls{zno} sample to be \qty{235}{\pico\second} and a rise time of \qty{20}{\pico\second} for alpha particle induced \gls{tcspc}. As \gls{zno} has been known for some time for its potential in fast timing applications, growth techniques, and different dopants have been investigated, e.g., by Neal et al. in \cite{neal2008zno_dopant_evaluation} and by \citet{bourret2009development}. 
\\

\subsection{Light Yield}
\citet{yanagida2021photoluminescence_Gan} report a light yield of \qty{5300}{\photons} for a \qty{5.5}{\MeV} alpha particle for the fast direct bandgap emission of \gls{gan} at $\approx$\qty{400}{\nano\meter}. For \gls{zno}, \citet{cates2013zno_ga} measured less than \qty{750}{\ph/\mega\electronvolt}$_{\alpha}$ from a powder sample, where \qty{}{\ph/\mega\electronvolt}$_{\alpha}$ describes the light yield per alpha particle energy. \gls{yap} is a scintillator with a well-known light yield. Crytur, the manufacturer, reports a light yield of \linebreak[4] \qty{5630\pm630}{\ph/\mega\electronvolt}$_{\alpha}$, where the error arises from the uncertainty in the alpha-to-gamma ratio, ranging from \numrange{0.20}{0.25}.

\subsection{Statistical Timing Performance Limits}
\label{sec:timing_models}
\citet{cates2013zno_ga} demonstrated sub-\qty{100}{\pico\second} timing at modest detected-photon counts for a \gls{zno} powder sample and benchmarked results against a straight-response model and the statistical \gls{crlb}. 
\\

A useful way to organize timing analysis for scintillators is to separate (i) the scintillation emission kinetics from (ii) the photodetector time response and the time pickoff. Following \citet{seifert2012crlb}, we use a straight-response model in which the stochastic emission process is convolved with the Gaussian sensor \gls{tts}, and is then connected to a leading-edge pickoff on the earliest part of the rising waveform. Full derivations and additional background are given in \citep[pp.~514--520]{degroot2012probability} as well as in later lower-bound analyses \cite{cates2013zno_ga}.
\\

The emission \gls{p.d.f.}, conditioned on photon detection, is denoted \(\phi(t)\). For many materials, the early light is well described by a linear combination of bi-exponential cascades. Writing $p_k$ for the weight of the $k$-th cascade with rise and decay constants $\tau_{r,k}$ and $\tau_{d,k}$, a convenient normalized, causal form is
\begin{equation}
\label{eq:cates1}
\begin{aligned}
\phi(t) &= \sum_{k} p_k \,\frac{e^{-t/\tau_{d,k}} - e^{-t/\tau_{r,k}}}{\tau_{d,k}-\tau_{r,k}},\ \ \ t\ge0 \\
&\text{with } p_k \ge 0,\quad \sum_k p_k = 1 \, .
\end{aligned}
\end{equation} 
Convolving \(\phi(t)\) with a normal \gls{p.d.f.}\ \(g_\sigma(t)=\mathcal{N}(0,\sigma_{\mathrm{TTS}})\) that describes the photodetector's carrier \gls{tts} produces the \gls{p.d.f.} for the primary trigger times (conditioned on detection), shifted by the interaction time \(\theta\),
\begin{equation}
\label{eq:cates2}
f(t\mid\theta)\;=\;(\phi * g_\sigma)(t-\theta)\;=\;\int_{0}^{\infty}\!\phi(\tau)\,g_\sigma\!\bigl(t-\theta-\tau\bigr)\,d\tau\,,
\end{equation}
which we use as the likelihood for estimating \(\theta\) (the interaction time). In the straight-response picture, the expected analog waveform at the sensor input is obtained by filtering \(f(t\mid\theta)\) with the (normalized) single-photon response \(h(t)\), and its local slope at a chosen threshold predicts the leading-edge timing jitter. See \citet{cates2013zno_ga} for details of Eqs.\,(1)--(5).
\\

If \(N\) photons are detected and their trigger times are treated as independent and identically distributed samples from \(f(t\mid\theta)\), the Fisher information about \(\theta\) is
\begin{equation}
\label{eq:cates6}
\mathcal{I}(\theta)\;=\;N\int_{-\infty}^{\infty}\!\frac{\bigl(\partial_\theta f(t\mid\theta)\bigr)^2}{f(t\mid\theta)}\,dt.
\end{equation}
The \acrfull{crlb} on any unbiased time estimator \(\hat\theta\) is
\begin{equation}
\label{eq:cates7}\operatorname{Var}(\hat\theta)\;\ge\;\frac{1}{\mathcal{I}(\theta)}.
\end{equation}
In timing applications, one usually reports a \gls{fwhm}, so the corresponding resolution is
\begin{equation}
\label{eq:fwhm_fisher}
\mathrm{FWHM}_{\text{CRLB}}\approx 2\sqrt{2\ln 2}\,\sqrt{\operatorname{Var}(\hat\theta)}\,.
\end{equation}
To expose the dominant scaling with scintillation kinetics, consider the single–component limit of Eq.~\eqref{eq:cates1} with negligible rise time ($\tau_r \ll \tau_d$) and negligible sensor $\mathrm{TTS}$, such that $f(t\mid\theta) \approx \frac{1}{\tau_d} \exp\!\left(-\frac{t-\theta}{\tau_d}\right)\Theta(t-\theta)$. In this regime, $\partial_\theta f = \frac{1}{\tau_d} f$, and inserting into Eq.~\eqref{eq:cates6} yields
\begin{equation}
\mathcal{I}(\theta)
= N \int \frac{(\partial_\theta f)^2}{f}\,dt
= N \int \frac{1}{\tau_d^2} f(t\mid\theta)\,dt
= \frac{N}{\tau_d^2}.
\end{equation}
Using Eq.~\eqref{eq:cates7}, the variance therefore satisfies
\begin{equation}
\operatorname{Var}(\hat\theta) \ge \frac{\tau_d^2}{N},
\end{equation}
and consequently
\begin{equation}
\mathrm{FWHM}_{\text{CRLB}} \propto \frac{\tau_d}{\sqrt{N}}.
\end{equation}
Thus, once the rise time and detector jitter no longer dominate, the achievable timing resolution scales linearly with the decay constant and inversely with the square root of the detected-photon statistics.
\citet{cates2013zno_ga} show that for well–conditioned signals (low noise, high gain) and with an optimized leading–edge pickoff on the earliest rise, measured resolutions can closely approach this statistical limit.

\subsection{Position Resolution}
\citet{ayllon2021_api_3d_recon} characterized the position resolution capabilities of a \gls{yap} scintillator in an integrated \gls{api} system and reported an achievable alpha particle resolution using a Hamamatsu H13700-03 SEL \gls{pmt} of less than \qty{0.2}{\milli\meter}. For \gls{zno}, \citet{cates_zno_pos_res} report a position resolution of less than \qty{1}{\milli\meter} for \qty{80}{\percent} of the detector area.

\FloatBarrier
\section{Samples studied in this Work}
\label{sec:samples}
\begin{figure}
  \centering

  \begin{subfigure}{0.8\linewidth}
    \centering
    \includegraphics[width=\linewidth]{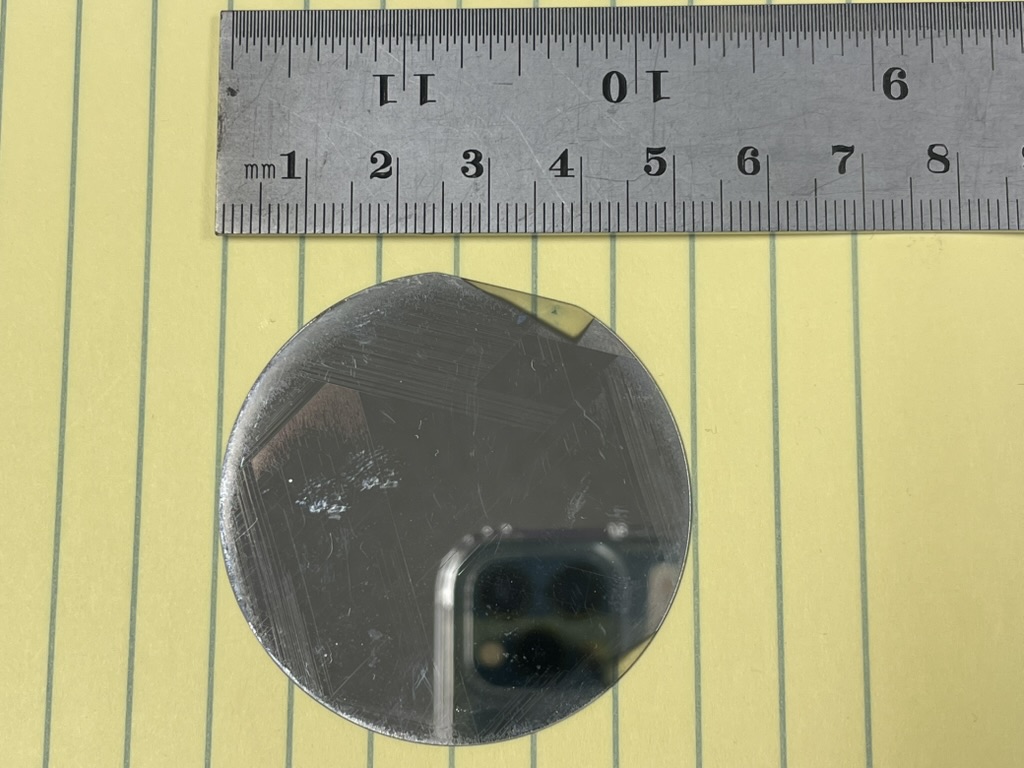}
    \caption[GaN sample]{\texorpdfstring{\gls{gan}}{GaN} sample: \qty{10}{\micro\metre} \texorpdfstring{\gls{gan}}{GaN} at \qty{7e18}{\per\cubic\centi\metre}, with a \qty{250}{\nano\metre} Al coating, grown epitaxially on a \qty{50}{\milli\meter} diameter, \qty{430}{\micro\metre} thick sapphire wafer.}
    \label{fig:gan_sample}
  \end{subfigure}

  \begin{subfigure}{0.8\linewidth}
    \centering
    \includegraphics[width=\linewidth]{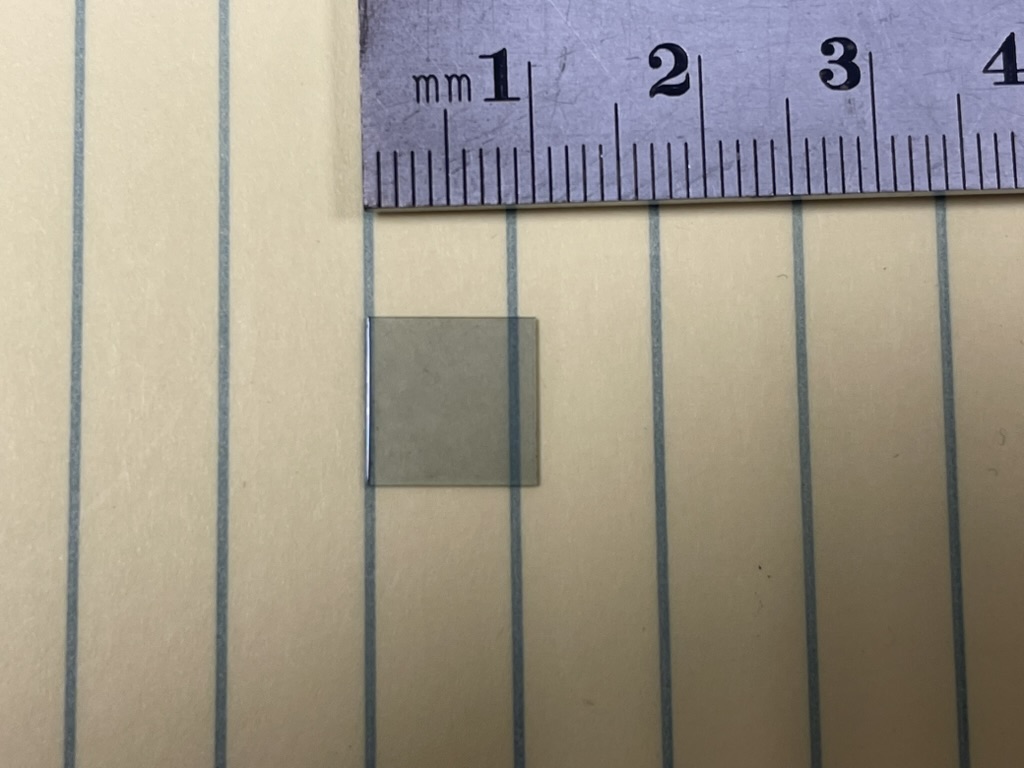}
    \caption[ZnO sample]{\texorpdfstring{\gls{zno}}{ZnO} sample: \qtyproduct{10 x 10 x 0.5}{\milli\metre} with a Gallium concentration of 1-3\%. The sample was provided by Gamma Reality Inc. and sourced from the MTI Corporation.}
    \label{fig:zno_sample}
  \end{subfigure}

  \begin{subfigure}{0.8\linewidth}
    \centering
    \includegraphics[width=\linewidth]{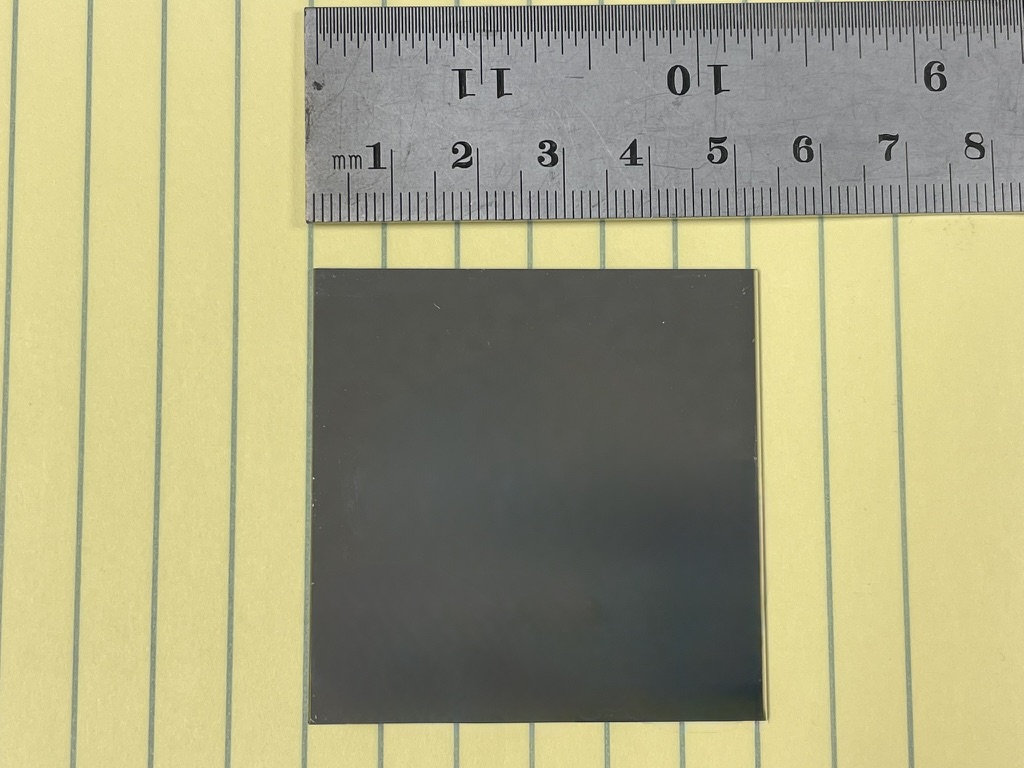}
    \caption[YAP:Ce sample]{\texorpdfstring{\gls{yap}}{YAP:Ce} sample: \qtyproduct{50 x 50 x 1}{\milli\metre} 
      with \qty{400}{\nano\metre} Al coating. The sample was grown by Crytur.}
    \label{fig:yap_sample}
  \end{subfigure}

  \caption[Samples used]{\texorpdfstring{\gls{gan}}{GaN}, \texorpdfstring{\gls{zno}}{ZnO}, and 
    \texorpdfstring{\gls{yap}}{YAP:Ce} samples used in this work.}
  \label{fig:samples}
\end{figure}

Three different scintillator samples were studied in this work and are shown in \autoref{fig:samples}. The \gls{gan} sample consisted of a \qty{50}{\milli\metre}-diameter, \qty{430}{\micro\metre}-thick sapphire wafer bearing epitaxially grown layers: \qty{1.5}{\micro\metre} of unintentionally doped GaN and a \qty{10}{\micro\metre} silicon-doped GaN film with a silicon concentration of \qty{7e18}{\per\cubic\centi\metre}. An additional \qty{250}{\nano\metre} Al coating was deposited on the top surface to enhance light-collection efficiency. The structure was grown by \gls{movpe} and provided by the group of Michał Boćkowski (Institute of High-Pressure Physics, Polish Academy of Sciences, Warsaw).

The \gls{zno} sample was a bulk single-crystal with lateral dimensions of \qtyproduct{10 x 10}{\milli\metre} and a thickness of \qty{0.5}{\milli\metre}. One face was polished. The crystal was doped with gallium at a concentration of $1-3\%$, purchased from MTI Corporation, and supplied by Gamma Reality Inc.

Finally, the \gls{yap} sample consisted of a \linebreak[4] \qtyproduct{50 x 50 x 1}{\mm} \gls{yap} single-crystal grown by Crytur. A \qty{400}{\nano\metre} Al coating was applied to the surface to improve light collection.

\subsection{Scintillation Properties}
The luminescence of the wide-bandgap semiconductor \gls{gan} is governed by a competition between near-band-edge recombination and defect-related emission bands. At room temperature, the direct bandgap of GaN is near \qty{3.4}{\electronvolt}, corresponding to ultraviolet emission around \qty{365}{\nano\metre} (see, for example, \citet{pittet2009_doping_effect_Si_in_gan}). For heavily Si-doped material, the radiative efficiency increases due to enhanced band-to-band recombination rates \cite{schenk2008gan_si_doping}. In addition to band-edge emission, a broad yellow luminescence band centered around \qty{540}{\nano\metre} is typically observed. This band is commonly attributed to deep defect complexes such as Ga vacancies combined with oxygen or carbon impurities, which act as deep acceptors. 
\\

\gls{zno} is another wide-bandgap semiconductor with a band edge around \qty{3.37}{\electronvolt}. The dominant radiative mechanism involves recombination of electrons from Ga$^{3+}$ donor levels with ionized holes trapped at shallow acceptors, most effectively hydrogen ions bound to zinc vacancies. \citet{bourret2009development} have shown that processing ZnO with Ga$_2$O$_3$ followed by hydrogen annealing minimizes defects, resulting in high luminosity and stable sub-nanosecond decay. This donor–acceptor channel produces bright emission centered near \qty{389}{\nano\metre}, slightly below the band edge by about \qty{0.18}{\electronvolt}, and explains the ultra-fast, spectrally pure scintillation observed in an optimally prepared sample. In this study, the emission peak of the \gls{zno} sample is at \qty{401}{\nano\meter}, which suggests strong self-absorption in our relatively thick sample for the \qty{389}{\nano\meter} emission peak reported by \citet{bourret2009development}. Therefore, unlike \gls{gan}, high-quality \gls{zno} shows little contribution from deep-level luminescence, and its performance is set by the balance between Ga donor electrons and shallow acceptors created during processing.
\\

\subsection{Self-Absorption in \gls{gan} and \gls{zno}} 
\label{subsec:self-absorption}
Both \gls{gan} and \gls{zno} exhibit strong self-absorption of their emitted light because the emission lies close to the band gap --- only about \qty{0.07}{\electronvolt} below the band edge in \gls{gan} and \qty{0.18}{\electronvolt} below the band edge in \gls{zno}. The Urbach tail describes absorption near the band edge as a function of the energy difference between the emission and the gap, $E_{\text{em}} - E_g$ \cite{cremades2000GaN_Si_absorption_doping_carrier_concentration}:
\begin{equation}
  \alpha(E_\text{em}) = \alpha_0\,\exp\!\left(\frac{E_{\text{em}} - E_g}{E_U}\right),
  \label{eq:urbach}
\end{equation}
where $\alpha(E_\text{em})$ is the absorption coefficient at the emission energy, $\alpha_0$ is the band-edge absorption coefficient, and $E_U$ is the Urbach energy, a measure of the tail width. In heavily doped \gls{gan}, for example, a doping level of about \qty{8e18}{\centi\metre^{-3}} may yield \(E_{U\text{, \gls{gan}}} \approx \qty{50}{\milli\electronvolt}\) \cite{cremades2000GaN_Si_absorption_doping_carrier_concentration} and \(E_{U\text{, \gls{zno}}} \approx \qty{90}{\milli\electronvolt}\) for \gls{zno} \cite{yilmaz2015ZnO_Ga_Urbach_tail}. Because the emission energy lies only slightly below the band gap in both materials, the absorption coefficient remains high in the emission region: evaluating $\alpha$ at $E=E_{\text{em}}$ gives $\alpha(E_{\text{em}})/\alpha_0 \approx 14\%$ for \gls{zno} and \(\approx 25\%\) for \gls{gan}. At the band-edge energy, the absorption coefficients are \(\alpha_0 = \qty{1e4}{\centi\metre^{-1}}\) for \gls{gan} and \(\alpha_0 = \qty{1.6e4}{\centi\metre^{-1}}\) for \gls{zno} \cite{schenk2009cathodoluminescence_gan_si}, which yield \(\alpha(E_{\text{em}}) \approx \qty{2470}{\centi\metre^{-1}}\) for \gls{gan} and \(\approx \qty{2160}{\centi\metre^{-1}}\) for \gls{zno}. In contrast, in a host such as \gls{yap} with a much larger band gap of \qty{8.1}{\electronvolt} \cite{yu2015energy}, the dopant emission at \qty{370}{\nano\meter} lies far below the absorption edge and is transmitted with negligible reabsorption.

\section{Experimental Methods}
\label{sec:experimental_methods}
Rise and decay times are obtained via X-ray-induced \gls{tcspc}, which probes the intrinsic scintillation kinetics of each material and provides results that are broadly applicable regardless of the excitation particle type. Light yield, \gls{ctr}, and position resolution are measured under direct alpha particle irradiation, making the results specific to alpha detection applications such as \gls{api}. The timing framework and \gls{crlb} analysis, however, are general and applicable to broader contexts, including \gls{pet} and \gls{ct}.
\subsection{Time-Correlated Single-Photon Counting Setup}
\label{subsec:tcspc_setup}
\begin{figure}
    \centering
    \includegraphics[width=0.5\textwidth]{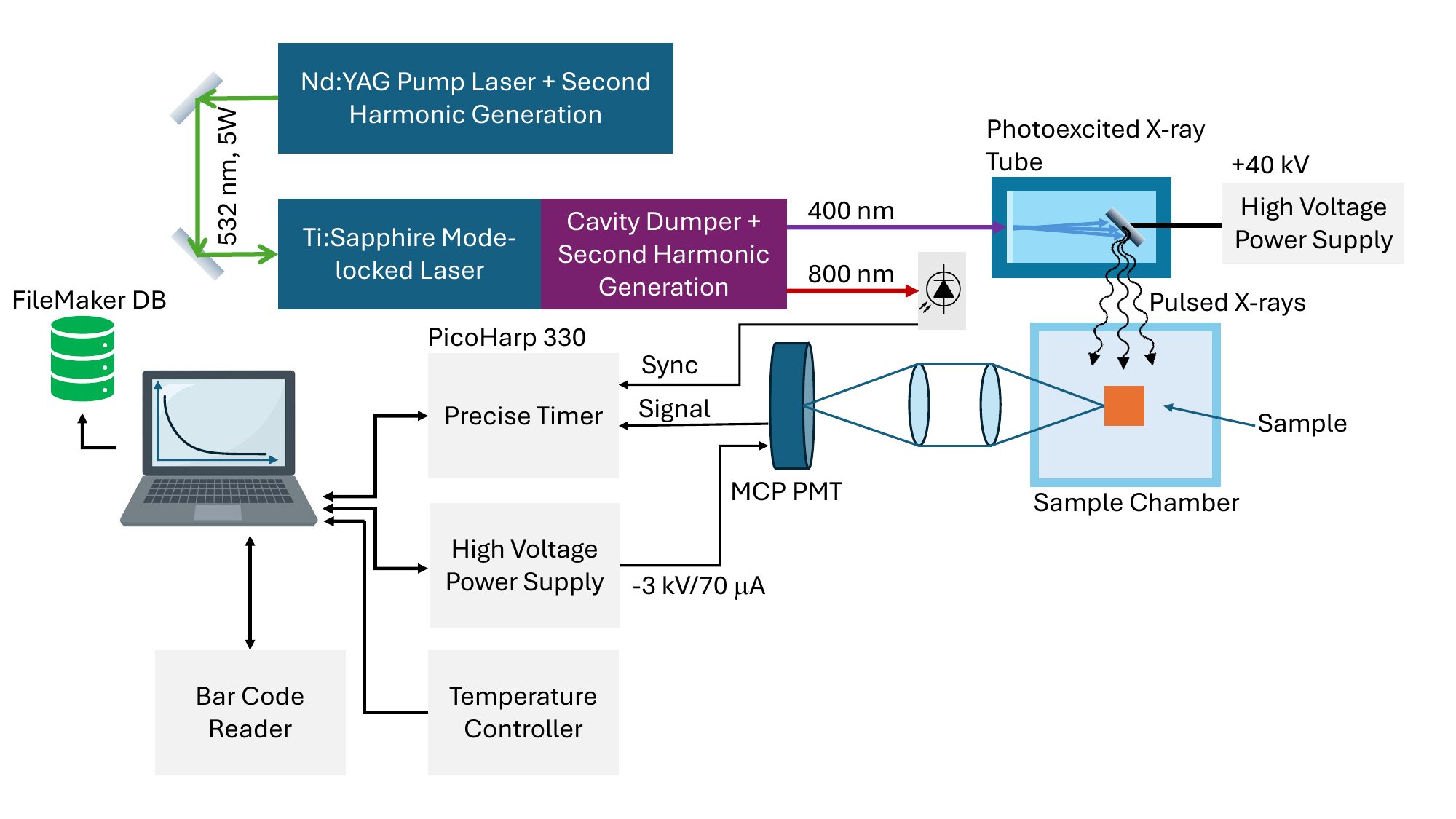}
    \caption{\gls{tcspc} setup used to measure scintillation rise and decay times. A pulsed Ti:Sapphire laser in 2nd harmonic mode drives a photo-excited X-ray tube, and scintillation photons are detected by an \gls{mcp}.}
    \label{fig:tcspc_setup}
\end{figure}

To quantitatively characterize the scintillation kinetics of the different samples and to calculate the \gls{crlb} according to Eq.~\eqref{eq:fwhm_fisher}, \acrlong{tcspc} measurements were performed. The experimental setup, shown in \autoref{fig:tcspc_setup}, employed an ultra-short pulsed \qty{200}{\femto\second} Ti:Sapphire laser with frequency-doubled output (via second harmonic generation), pumped by a Nd:YAG laser operating at its second harmonic. The light from the Ti:Sapphire laser served both as the start reference and as the driver for a photo-excited X-ray tube (Hamamatsu N5084) operating at \qty{40}{\kilo\volt}. The emitted X-rays excited scintillation light in the samples, which was collected by a Hamamatsu R3809U-50 \gls{mcp} with an intrinsic \gls{tts} of \qty{25}{\pico\second} \gls{fwhm}. Photon arrival times were digitized with a PicoHarp 330, which uses \gls{cfd} for accurate time-stamping of the start and stop signals and records their time differences. An older version of this setup is described by \citet{derenzo1992design_tcspc}. The \gls{irf} of the system, assumed to be Gaussian, has been improved since the original publication of the design from \qty{109}{\pico\second} to about \qty{68}{\pico\second} \gls{fwhm}. 
The \gls{irf} of the system was measured using a PbF$_2$ sample. While the IRF is not perfectly Gaussian, a Gaussian fit was used to estimate its \gls{fwhm}. Since the original publication of the design, the IRF has been improved from \qty{109}{\pico\second} to about \qty{68}{\pico\second} \gls{fwhm}. The timing resolution is predominantly limited by the response of the pulsed X-ray tube, whereas contributions from the PMT transit time spread (\(\sim\) \qty{25}{\pico\second}) and the excitation pulse duration (\(\sim\) \qty{200}{\femto\second}) are comparatively small.

\subsection{Light Yield Measurement Setup}
\label{subsec:ly_setup}
\begin{figure}
    \centering 
    \includegraphics[width=0.48\textwidth]{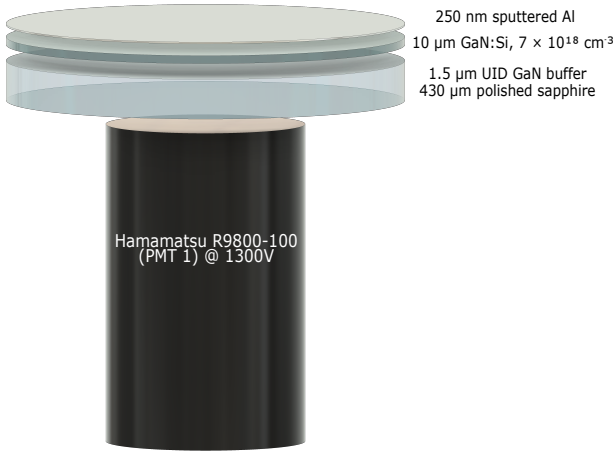}          
    \caption{Schematic drawing of the aluminized \gls{gan} scintillation layer on a Sapphire wafer, coupled to a R9800-100 Hamamatsu \gls{pmt} to measure the light yield.}\label{fig:GaN_on_PMT}
\end{figure}

To measure the absolute light yield of the \gls{gan}, \gls{zno}, and \gls{yap} samples in the setup shown in \autoref{fig:GaN_on_PMT}, the single-photon response of the Hamamatsu R9800-100 \gls{pmt} with the DRS4 evaluation board data acquisition system was first quantified. 
\subsubsection{Single-Photon Response Setup}
\label{subsubsec:single_photon_response_setup}
\begin{figure}
      \centering
      \includegraphics[width=\linewidth]{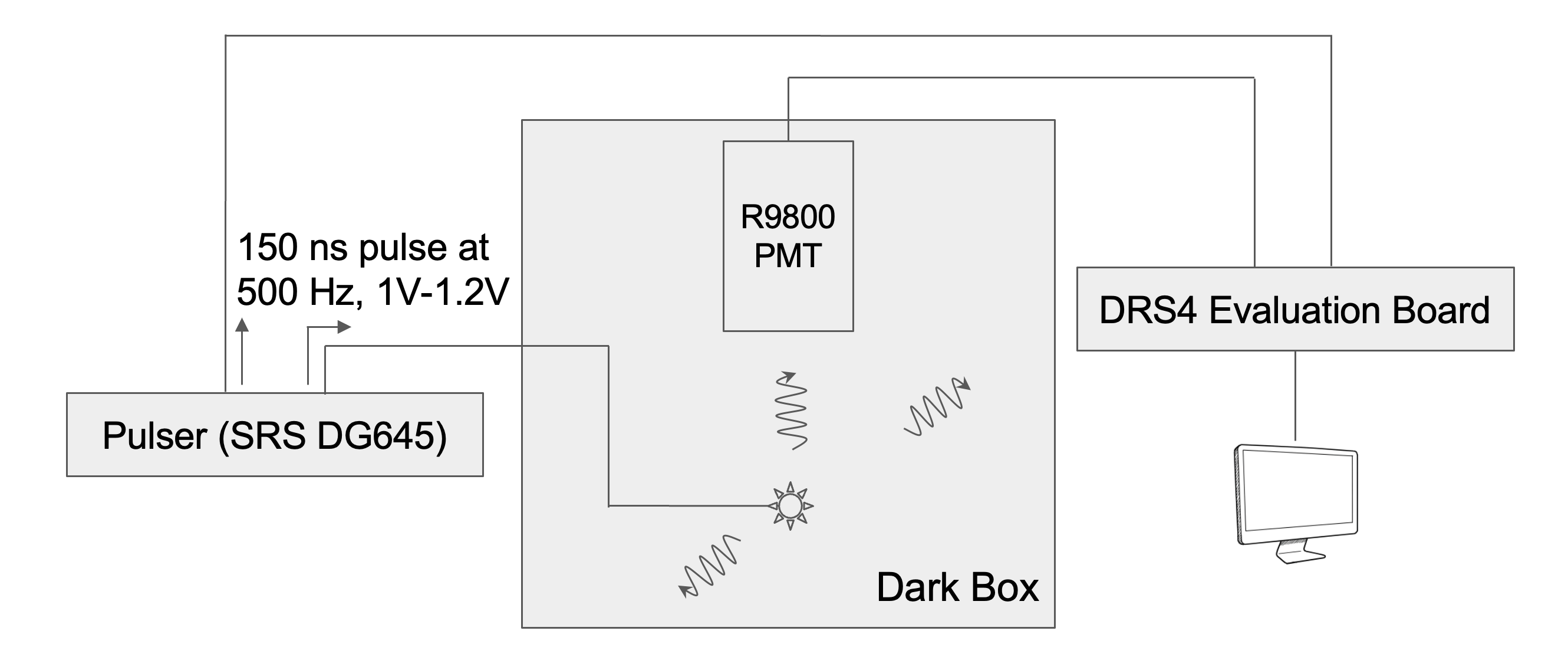}
      \caption{Experimental setup to measure the single-photon response.}
      \label{fig:spr_setup}
\end{figure}

The experimental setup to measure the single-photon response is shown in \autoref{fig:spr_setup}. An LED was pulsed at \qty{500}{\hertz} with a pulse duration of \qty{150}{\nano\second}. The same pulse was used to trigger the data acquisition system. Varying the applied LED voltage produced a shift in detected photons. 

\subsubsection{Photon Transport Efficiency Simulation}
\label{subsubsec:photon_transport_efficiency_simulation}
\begin{figure}
    \centering 
    \includegraphics[width=0.48\textwidth]{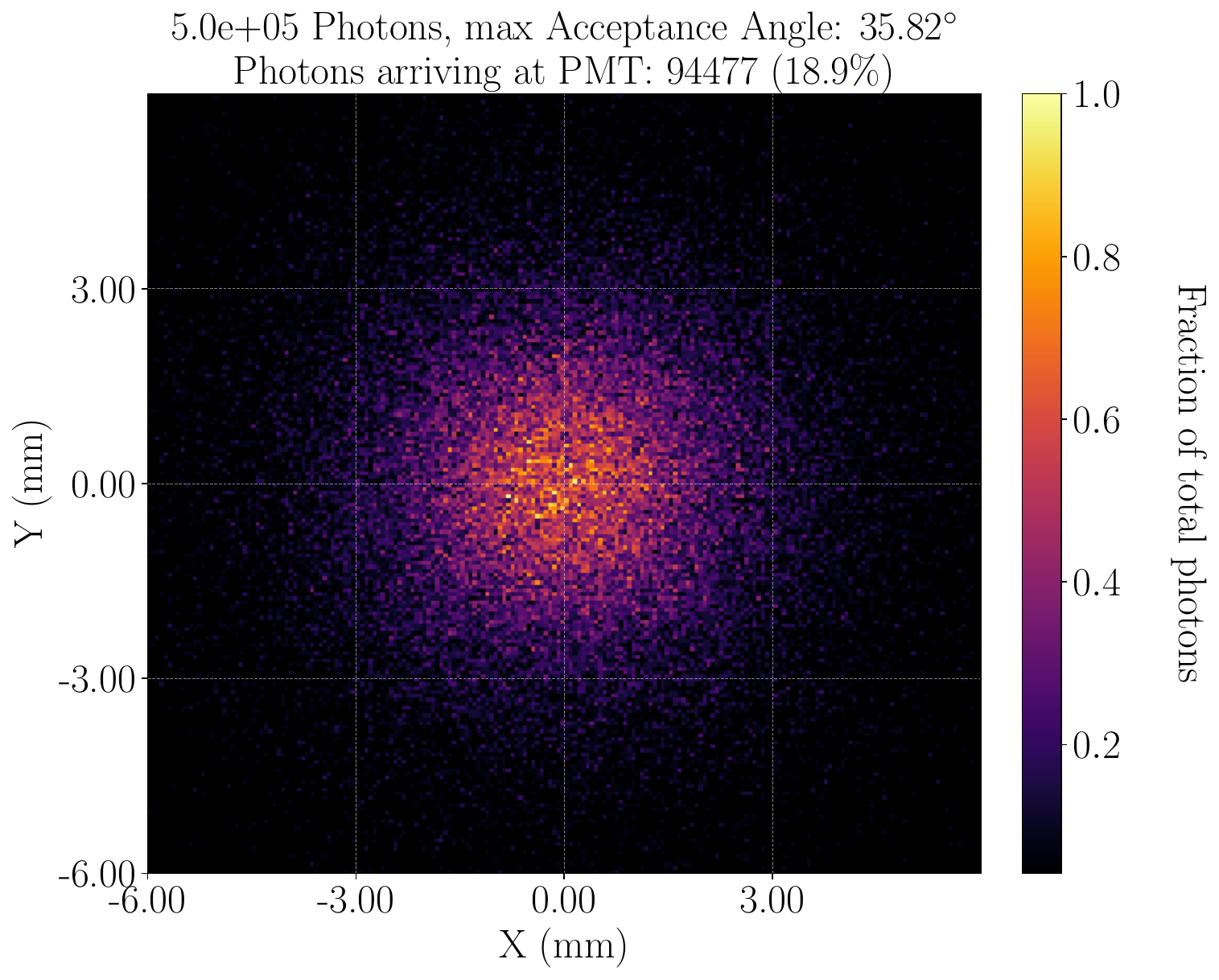}          
    \caption{Monte Carlo photon transport simulation for \gls{am241} induced photons created in \gls{gan}. The distribution shows the fraction of \qty{18.9}{\percent} of photons arriving at the \gls{pmt} window for alpha particles hitting in the origin of the plot.}\label{fig:GaN_photon_transport_simulation}
\end{figure}

To derive the number of photons created by a charged particle in a scintillator from the number of detected photoelectrons (p.e.), one must account for the photon-transport efficiency from the scintillator into the PMT. The geometry for this is shown for the \gls{gan} sample in \autoref{fig:GaN_on_PMT}. A similar geometry was used for the \gls{zno} and \gls{yap} samples. Using a Monte Carlo photon transport simulation (see \autoref{fig:GaN_photon_transport_simulation}), the photon-transport efficiency is defined as the fraction of photons arriving at the \gls{pmt}. For \gls{gan}, \gls{zno} and \gls{yap}, the photon transport efficiency was simulated to be \qty{18.9\pm0.2}{\percent}, \qty{23\pm0.2}{\percent} and \qty{34.5\pm0.3}{\percent}, respectively. The simulation assumes that all photons are created at a single point in space and time, which is a valid approximation for the bulk materials used in this study, where particle ranges are less than \qty{15}{\micro\meter} and stopping times are less than \qty{2}{\pico\second}. The simulation models absorption and transmission using the Beer–Lambert law and the Fresnel equations. The main reason for photon losses is the index of refraction mismatch. The values for indices of refraction, absorption coefficients, and the thickness of the relevant layers used in the simulation are summarized in \autoref{tab:optical_properties}. Note that only for the \gls{gan} sample there is a sapphire layer between the substrate and the optical grease to couple to the borosilicate glass \gls{pmt} window.
\begin{table}[htbp]
    \centering
    \caption{Comparison of scintillator materials and materials used in the simulation regarding their optical properties: Refractive index $n$, peak emission wavelength, absorption coefficient $\alpha$, and layer thickness. Data is obtained from \cite{refractive_indices_database} and \cite{schenk2009cathodoluminescence_gan_si} in combination with \autoref{subsec:self-absorption} to get the absorption values for \gls{gan} and \gls{zno} at the measured emission wavelengths.}
    \begin{tabular}{lcccc}
        \hline
        Material & $n$ & $\lambda_\mathrm{em}$ (\qty{}{\nano\meter}) & $\alpha$ (\qty{}{\centi\meter^{-1}}) & Layer (\qty{}{\meter})\\
        \hline
        YAP:Ce   & 1.931 & 370 & $\ll 1$ & $1 \cdot 10^{-3}$\\
        GaN:Si   & 2.505 & 372 & 2470 & $1.1 \cdot 10^{-5}$\\
        ZnO:Ga   & 1.706 & 401 & 2160 & $5 \cdot 10^{-4}$\\
        Sapphire & 1.793 & n.a. & $\ll 1$ & $5 \cdot 10^{-4}$\\
        Opt. grease & 1.466 & n.a. & $\ll 1$ & $2 \cdot 10^{-4}$\\
        PMT glass & 1.535 & n.a & $\ll 1$ & $1.5 \cdot 10^{-3}$\\
        \hline
    \end{tabular}
    \label{tab:optical_properties}
\end{table}

\FloatBarrier
\subsection{Coincidence Timing Resolution Setup}
\label{subsec:timing_setup}
\begin{figure}
  \centering
  \begin{subfigure}{\linewidth}
      \centering
      \includegraphics[width=0.9\textwidth]{GaN_3_PMT_setup_v2_annotated_cropped.pdf}
      \caption{\gls{ctr} geometry with three R9800-100 PMTs and \gls{am241} source. A thin EJ-214 reference film provides a fast start signal, the scintillator under test (\gls{gan}, \gls{zno}, or \gls{yap}) provides the stop signal. Distances and apertures were fixed across runs. While shown with some distance for better visibility, the source and scintillator were as close to the EJ-214 film as possible.}
      \label{fig:setup_3pmts_with_gan}
  \end{subfigure}

  \vspace{1em}

  \begin{subfigure}{\linewidth}
      \centering
      \includegraphics[width=\textwidth]{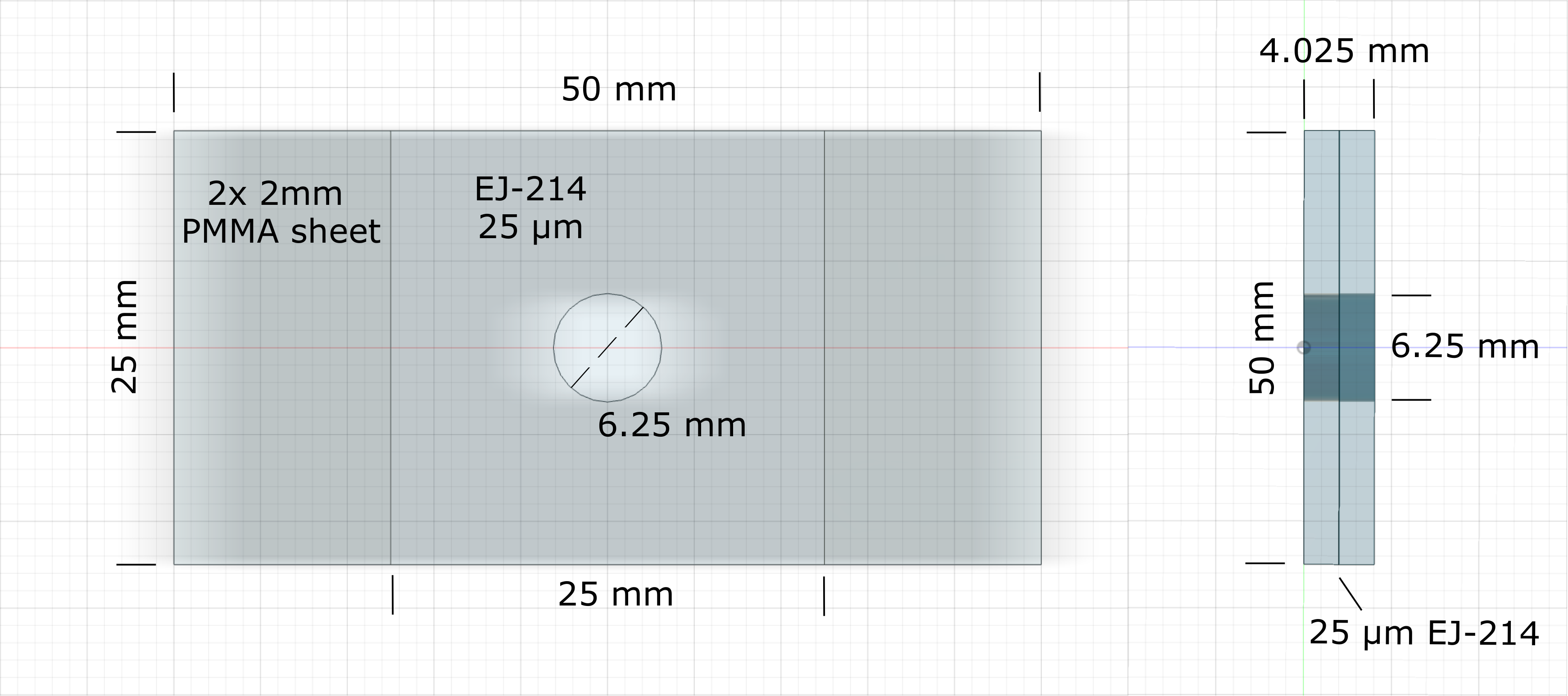}
      \caption{Thin film EJ-214 scintillator from Eljen Technology sandwiched between two \acrshort{pmma} light guides. Optical grease was used to fill the air gap between the \acrshort{pmma} slabs.}
      \label{fig:ej_from_top_annotated}
  \end{subfigure}

  \caption{Experimental setup for CTR measurements.}
  \label{fig:ctr_setup_combined}
\end{figure}

\autoref{fig:setup_3pmts_with_gan} shows the setup used to characterize the \gls{dtr} of the \gls{gan} sample. This setup is adapted from the setup developed by \citet{cates2013zno_ga}. The same configuration was used for \gls{yap} and \gls{zno}. Three Hamamatsu R9800-100 super bialkali-\glspl{pmt} were used to detect photons from the reference plastic scintillator and each sample scintillator. The \gls{tts} of \qty{115}{\pico\second} (\qty{270}{\pico\second} \gls{fwhm}) was taken from the Hamamatsu datasheet. A DRS4 evaluation board served as the data-acquisition system, offering \qty{700}{\mega\hertz} of analog bandwidth and a sampling rate of \qty{5e9}{\per\second} across four channels in coincidence. 

The alpha particles from a \qty{40}{\nano Ci} \qty{5.47}{\mega\electronvolt} \gls{am241} source, covered with a thin gold layer, first pass through a thin EJ-214 plastic scintillator (Eljen Technology) before they are stopped in the sample. \autoref{fig:ej_from_top_annotated} shows the EJ-214 film sandwiched between two \gls{pmma} light guides. The space between the \gls{pmma} slabs is filled with optical grease, and the assembly is wrapped in Teflon to increase reflectivity. The EJ-214-\gls{pmma} light guide structure is optically coupled to two Hamamatsu R9800-100 \glspl{pmt}, one on each side, so that a single alpha event is detected in both \glspl{pmt}. According to SRIM/TRIM simulations \cite{SRIM}, alpha particles deposit about \qty{2.9}{\mega\electronvolt} in the EJ-214 film and lose a total of about \qty{700}{\kilo\electronvolt} while traversing the \qty{4}{\milli\meter} of air gaps between the source, the EJ-214, and the scintillator, so that \qty{1.9}{\mega\electronvolt} are deposited in each scintillator sample.

\subsection{Position Resolution Measurement Setup}
\label{subsection:posresolution_exp_setup}
\begin{figure}
    \centering
    \includegraphics[width=\linewidth]{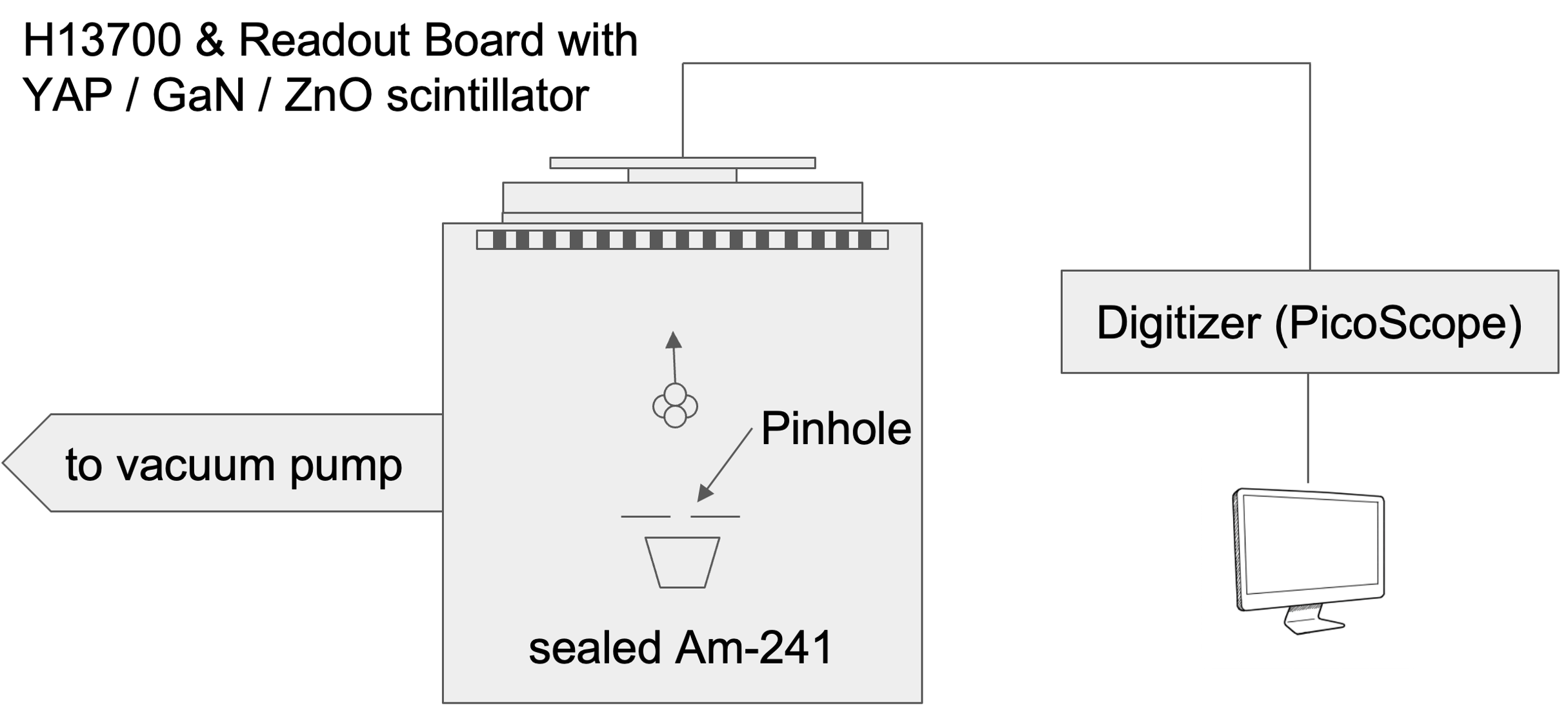}
    \caption{Experimental setup to measure the position resolution capabilities of all three scintillator samples in an emulated \gls{api} system.}
    \label{fig:pos_res_setup}
\end{figure}
\begin{figure}
    \centering
    \begin{subfigure}{0.47\linewidth}
        \centering
        \includegraphics[width=\linewidth]{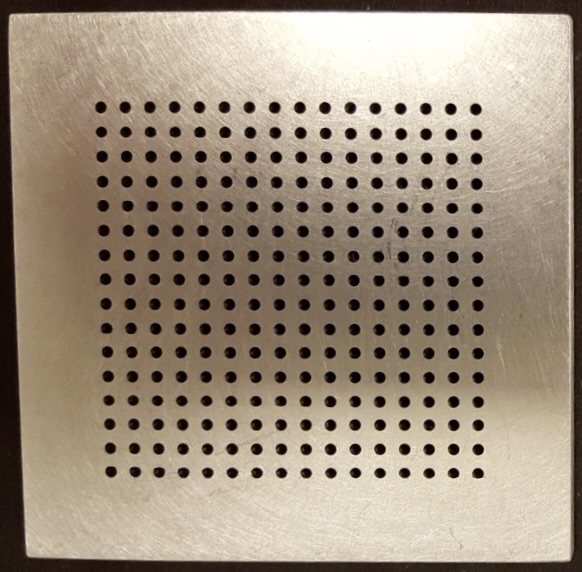}
        \caption{Coarse mask.}
        \label{fig:coarse_mask}
    \end{subfigure}
    \hfill
    \begin{subfigure}{0.45\linewidth}
        \centering
        \includegraphics[width=\linewidth]{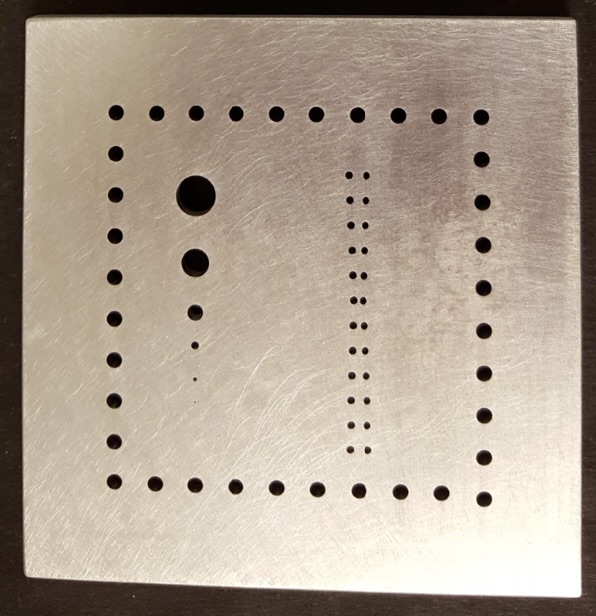}
        \caption{Fine mask.}
        \label{fig:fine_mask}
    \end{subfigure}
    \caption{Coarse and fine masks used in the position resolution measurements.}
    \label{fig:masks_combined}
\end{figure}

An \gls{am241} source was placed inside a vacuum chamber at a distance of about \qty{8}{\centi\meter} from the \gls{gan}, \gls{zno}, and \gls{yap} crystals. \autoref{fig:pos_res_setup} shows the experimental setup. 

The crystals were covered with the masks shown in \autoref{fig:masks_combined} to reconstruct the position resolution. Photons created in the crystals passed through a thin vacuum gap and a sapphire window, which served as the vacuum interface, into a Hamamatsu H13700-03 \gls{pmt} with \qtyproduct{16 x 16}{pixels} (\qtyproduct{3 x 3}{mm} pixel pitch) and a total active width of \qty{48}{\milli\meter}. This geometry is required not only to ensure that alpha particles have a line of sight to all parts of the \gls{pmt} but also to emulate the scintillator's implementation in an integrated \gls{api} system, as studied by \citet{ayllon2021_api_3d_recon}. 

The pixelated \gls{pmt} was read out with a custom board described in \cite{ayllon2021_api_3d_recon}. It uses a resistive network to encode the photon-impact position into four output channels. By integrating the signals on those four channels and computing the center of mass, a 2D interaction position is obtained.

\section{Results and Discussion}
\label{sec:results}
\subsection{Rise and Decay Time Results}
\label{subsec:tcspc_results}
\begin{figure}
    \centering
    \begin{subfigure}{0.44\textwidth}
        \centering
        \includegraphics[width=\textwidth]{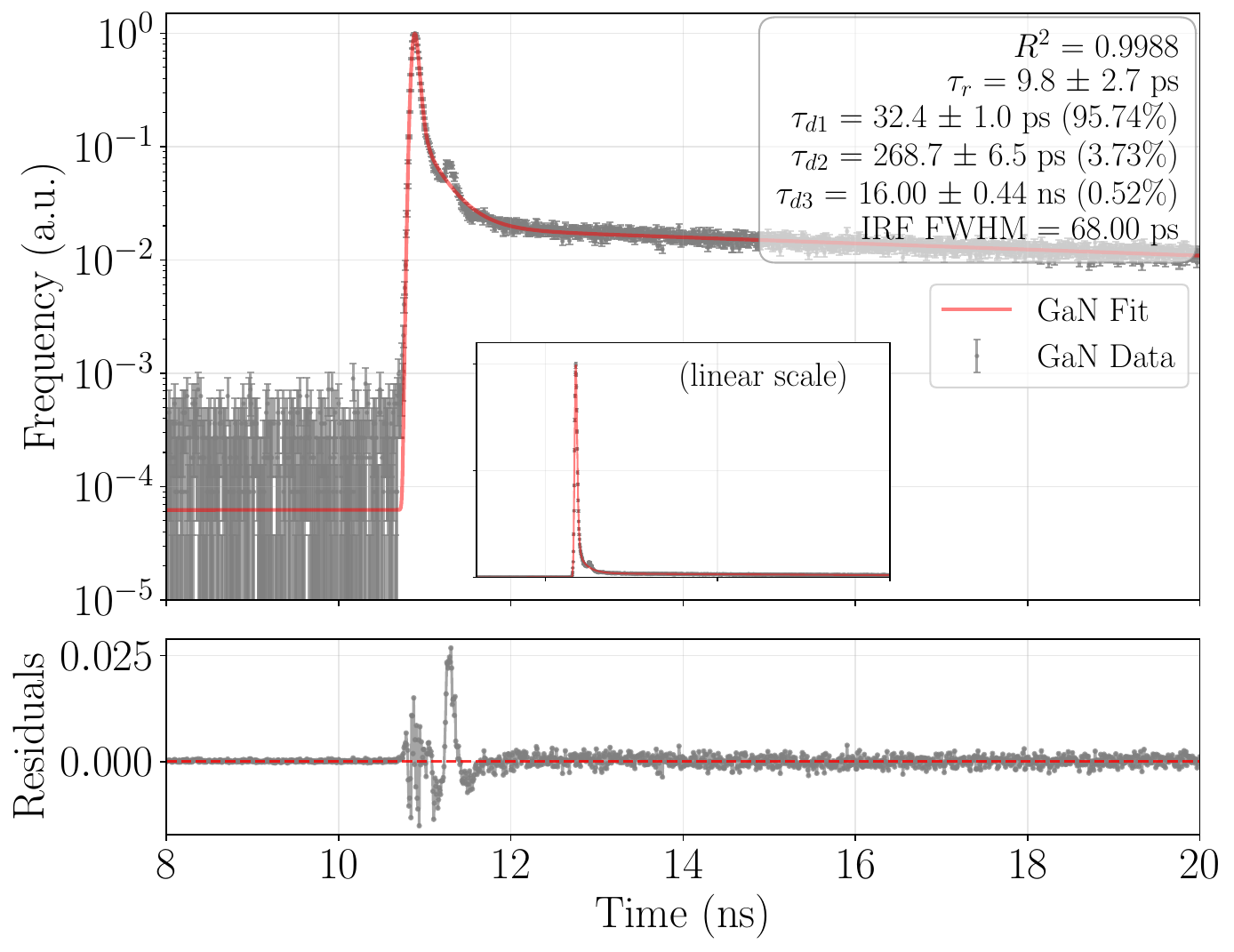}
        \caption{\gls{gan} sample.}
        \label{fig:GaN_TCSPC_trace}
    \end{subfigure}
    
        \vspace{0.5em}
    \begin{subfigure}{0.44\textwidth}
        \centering
        \includegraphics[width=\textwidth]{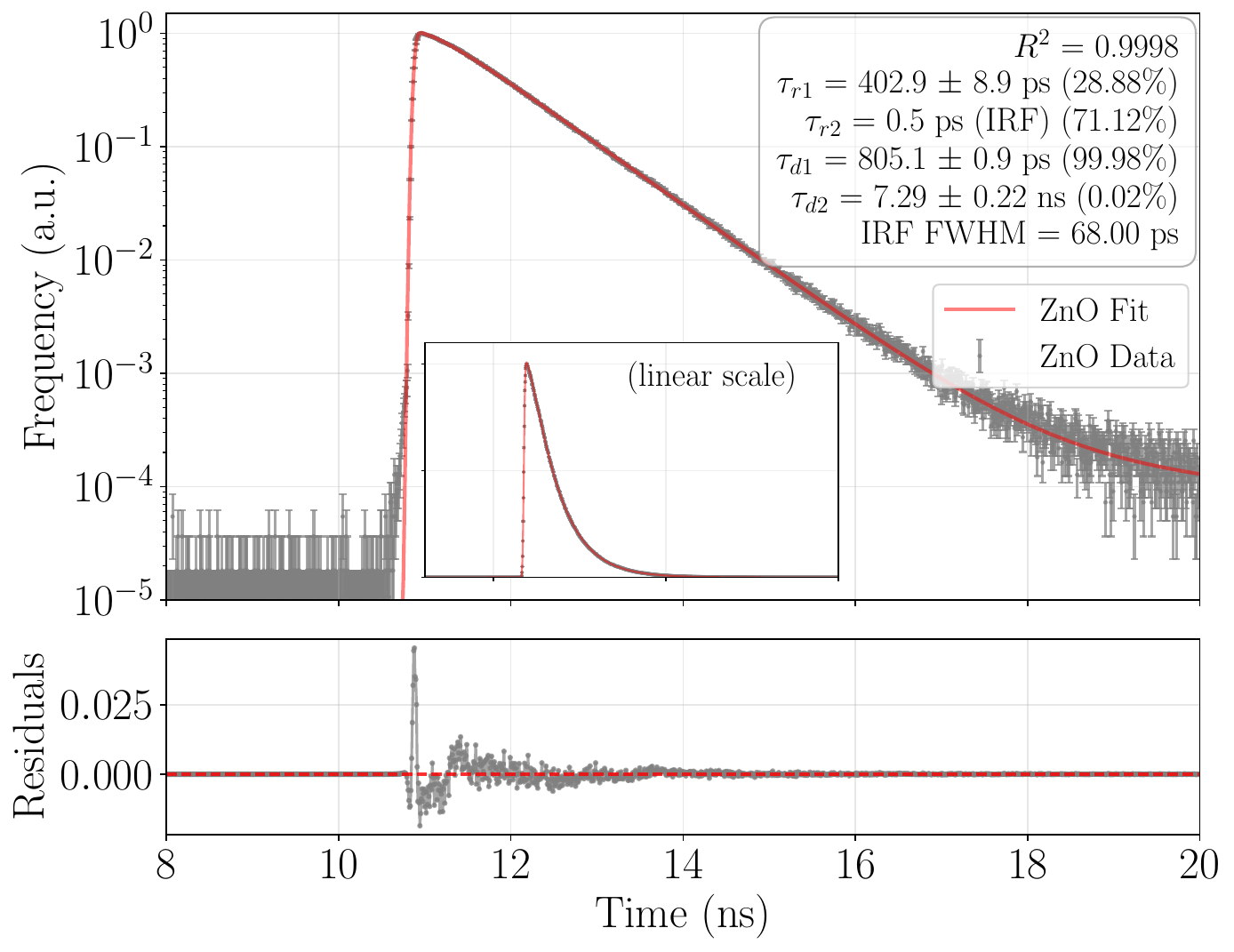}
        \caption{\gls{zno} sample.}
        \label{fig:ZnO_TCSPC_trace}
    \end{subfigure}
    
        \vspace{0.5em}
        
    \begin{subfigure}{0.44\textwidth}
        \centering
        \includegraphics[width=\textwidth]{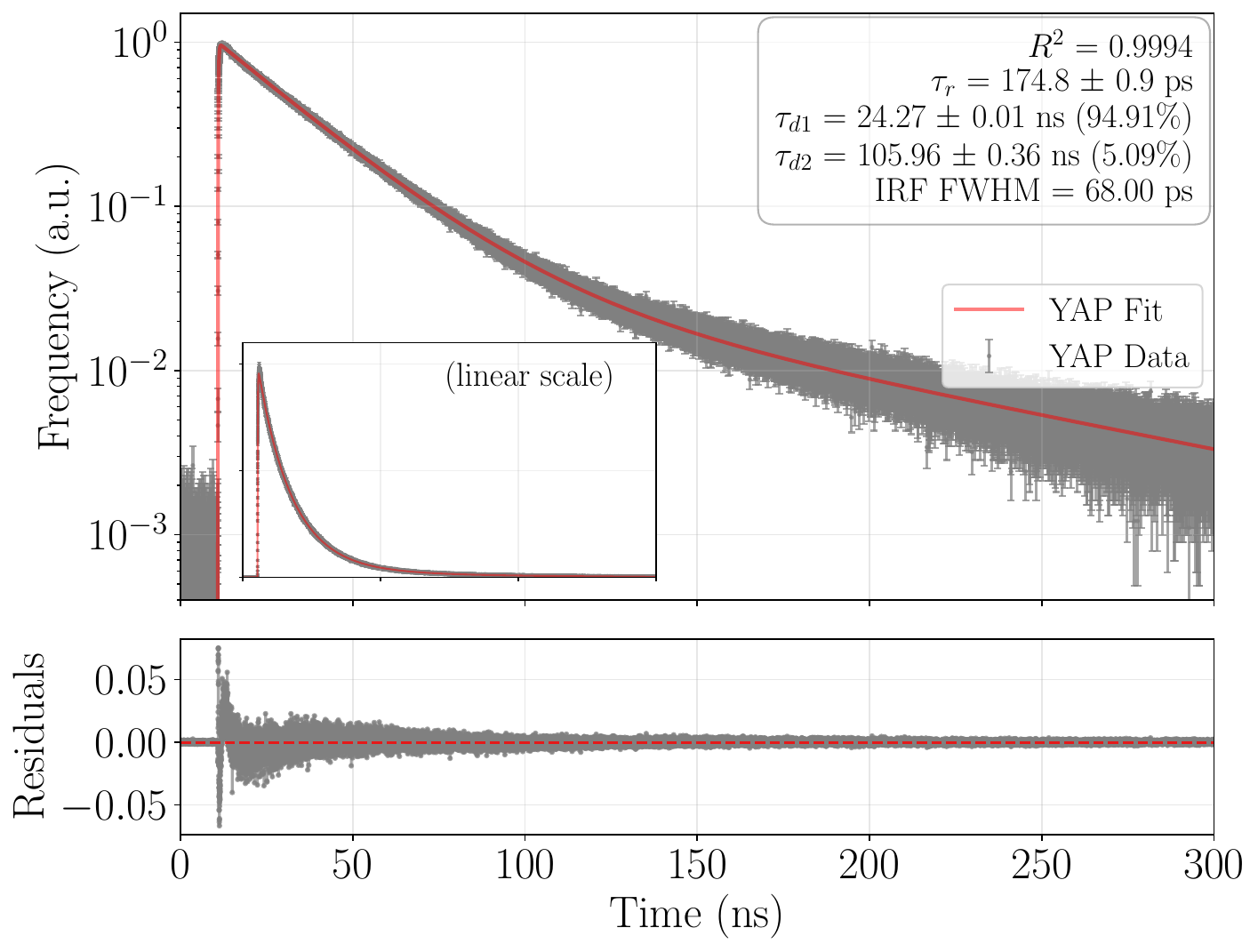}
        \caption{\gls{yap} sample. Note the different time scale.}
        \label{fig:YAP_TCSPC_trace}
    \end{subfigure}

    \caption{Measured \gls{tcspc} temporal distributions of the scintillation decay under pulsed X-ray excitation for (a) \gls{gan}, (b) \gls{zno}, and (c) \gls{yap} samples used in this work.}
    \label{fig:TCSPC_combined}
\end{figure}
Figs.~\ref{fig:GaN_TCSPC_trace}, \ref{fig:ZnO_TCSPC_trace}, and \ref{fig:YAP_TCSPC_trace} show the measured \gls{tcspc} histograms for \gls{gan}, \gls{zno}, and \gls{yap}. The insets show the same plots with a linear y-scale, and the fit residuals are shown below each \gls{tcspc} histogram. The waveform-shaped histograms reflect the convolution of the true scintillation decay with the \gls{irf}. Equation~\eqref{eq:cates1} was used as the fitting model, and a least-squares procedure was applied to extract characteristic rise and decay components. The fit results are summarized in \autoref{tab:rise_decay_comparison}. 
For \gls{yap}, the distribution shows a comparatively slow decay typical of Ce-activated oxides. By integrating a long (\qty{165}{\nano\second}) and a short (\qty{10}{\nano\second}) interval of the pulse, we obtain that about \qty{30}{\percent} of the total light yield is emitted in the first \qty{10}{\nano\second}. In contrast, \gls{zno} exhibits a sharp, sub-nanosecond peak consistent with its ultra-fast donor-acceptor recombination mechanism. We measured that about \qty{98}{\percent} of the total light yield is emitted in the first \qty{10}{\nano\second}. \gls{gan} shows intermediate behavior, with an ultra-fast near-band-edge component and an additional slower tail from defect-related recombination. Note that the slow component of \qty{16}{\nano\second} contributes only about \qty{0.5}{\percent} to the pulse shape. The small peak in the primary decay in \autoref{fig:GaN_TCSPC_trace} is likely related to instrumental artifacts. We measured that about \qty{80}{\percent} of the total light yield is emitted in the first \qty{10}{\nano\second}.

The quality of the fits and the agreement of the residuals confirm that the Gaussian \gls{irf} assumption is adequate to deconvolve the observed distributions. This analysis highlights the very different timing profiles of the three scintillators: \gls{zno} as an ultra-fast benchmark, \gls{yap} as a high-yield but slower reference, and \gls{gan} as a promising ultra-fast material that surpasses the well-established \gls{zno} in terms of timing and decay time but falls below \gls{zno} in light yield. 

  \begin{table}[htbp]
    \centering
    \small
    \setlength{\tabcolsep}{3pt}
    \caption{Measured rise and decay times for \gls{gan}, \gls{zno}, and \gls{yap} and their fractional weight in the fitted model}
    \label{tab:rise_decay_comparison}
    \begin{tabular}{lccc}
      \toprule
      \textbf{Material} &
      \textbf{Rise times} &
      \textbf{Decay times}\\
      \midrule
      GaN:Si &
      \qty{9.8 \pm 2.7}{\pico\second} &
      \makecell[l]{\qty{32.4\pm1.0}{\pico\second} (95.8\%) \\
                   \qty{269\pm7}{\pico\second} (3.7\%) \\
                   \qty{16.0\pm0.5}{\nano\second} (0.5\%)}\\
      \midrule
      ZnO:Ga &
      \makecell[l]{\qty{\le 5}{\pico\second} (71\%) \\
                   \qty{403\pm9}{\pico\second} (29\%)} &
      \makecell[l]{\qty{805\pm1}{\pico\second} (99.98\%) \\
                   \qty{7.3\pm0.3}{\nano\second} (0.02\%)} \\
      \midrule
      YAP:Ce &
      \qty{175\pm1}{\pico\second} &
      \makecell[l]{\qty{24.3\pm0.1}{\nano\second} (95\%) \\
                   \qty{106\pm0.4}{\nano\second} (5\%)}\\
      \bottomrule
    \end{tabular}
  \end{table}

\FloatBarrier
\subsection{Light Yield Results}
\label{subsec:ly_results}
\begin{figure}
      \centering
      \includegraphics[width=\linewidth]{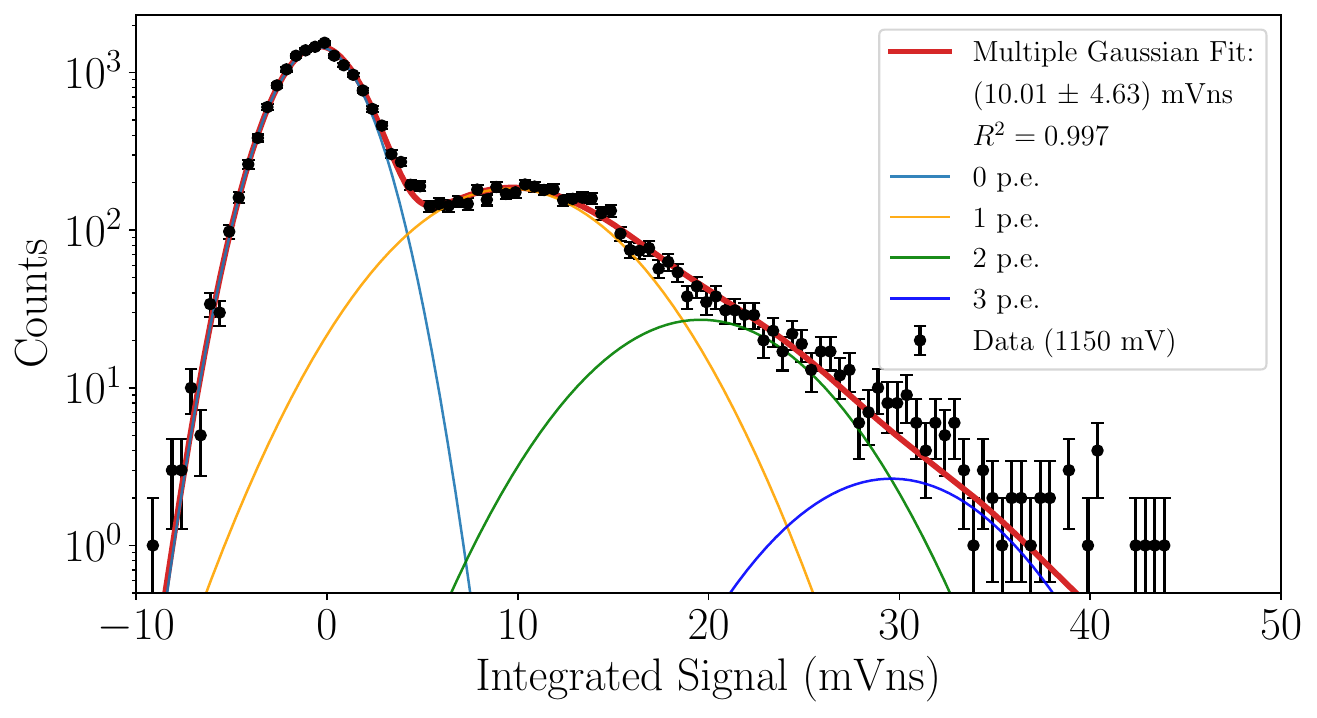}
      \caption{Integrated signal from the Hamamatsu R9800-100 \gls{pmt} when exposed to blue LED light at \qty{465}{\nano\meter} at \qty{1.15}{\volt}.}
      \label{fig:integrated_single_photons}
\end{figure}
\begin{figure}
    \centering
    \begin{subfigure}{\linewidth}
        \centering
        \includegraphics[width=\linewidth]{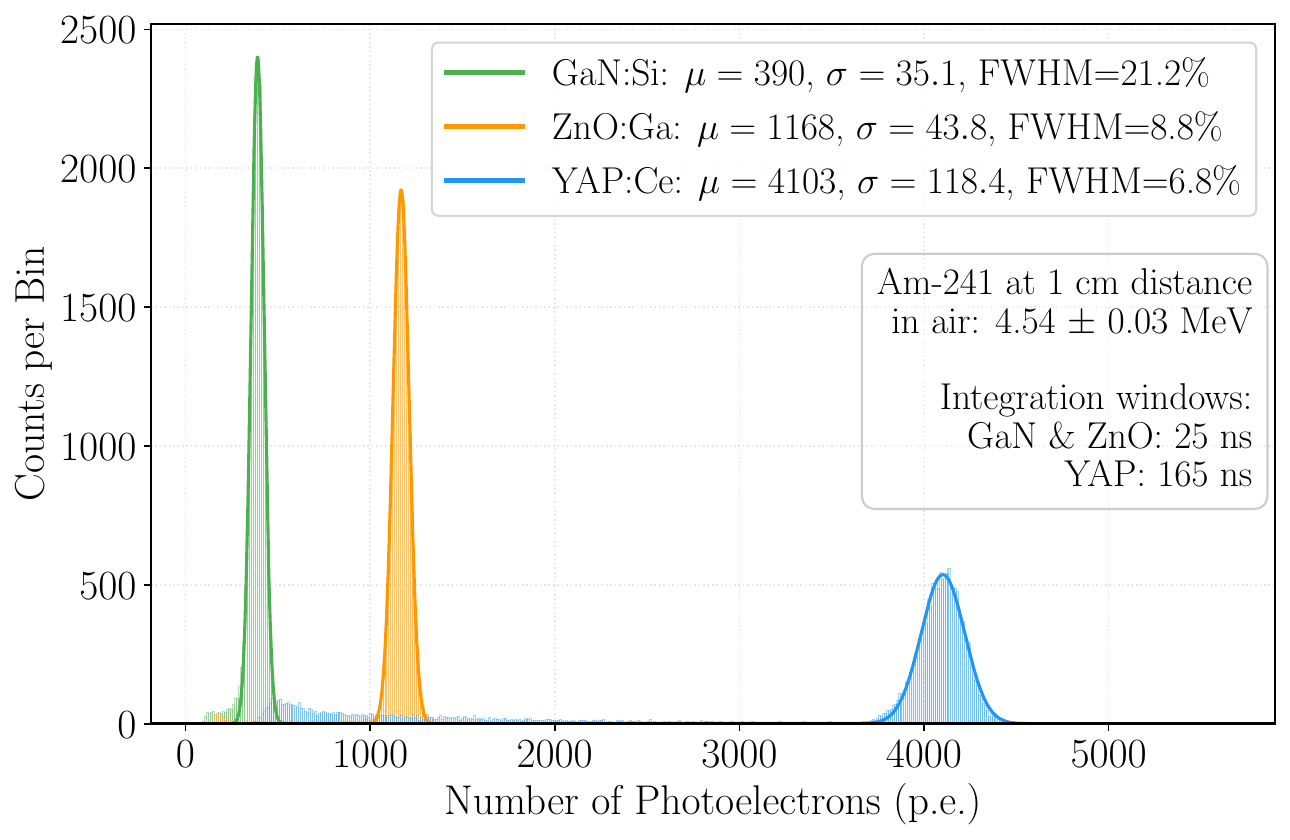}
        \caption{Effective light yield comparison for \gls{gan} (green), \gls{zno} (orange), and \gls{yap} (blue). Detected number of photoelectrons under \qty{5.47}{\mega\electronvolt} \gls{am241} at \qty{1}{\centi\meter} in air. Peaks: \gls{gan} $\mu=390$ p.e., \gls{zno} $\mu=1168$ p.e., \gls{yap} $\mu=4103$ p.e.}
        \label{fig:LY_comparison_detected_pe}
    \end{subfigure}
    
    \vspace{1em}
    
    \begin{subfigure}{\linewidth}
        \centering
        \includegraphics[width=\linewidth]{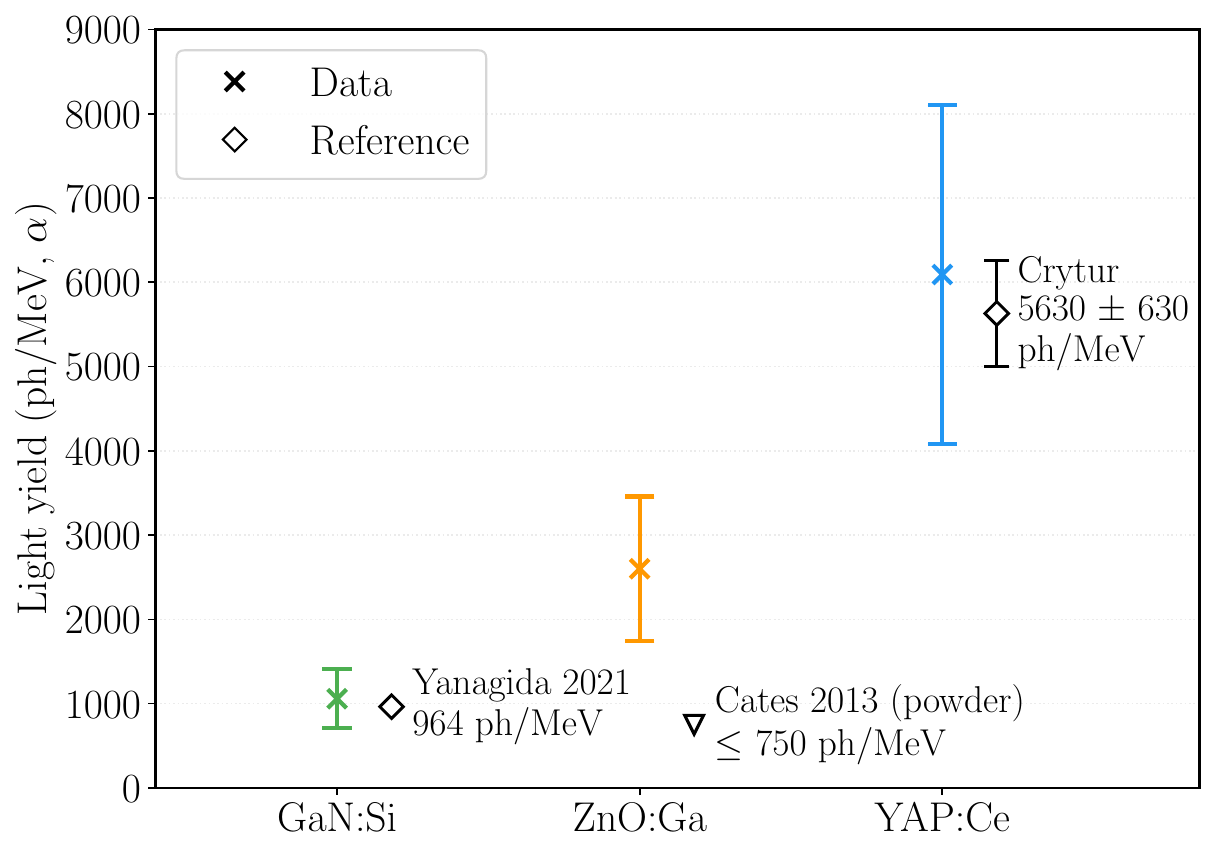}
        \caption{Absolute light yield comparison for \gls{gan}, \gls{zno}, and \gls{yap}. Created photons vs. literature.}
        \label{fig:LY_comparison_net}
    \end{subfigure}
    \caption{Experimentally detected and calculated absolute light yields for \gls{gan}, \gls{zno}, and \gls{yap} compared with literature values.}
    \label{fig:LY_combined}
\end{figure}

\subsubsection{Single-Photon Response Results}
To calculate the absolute light yield of the samples used in this work, both the single-photon response and the \gls{qe} of the \gls{pmt} must be known. We used Poisson statistics to extract the single-photon response from the pulsed LED setup discussed in \autoref{subsubsec:single_photon_response_setup}, as shown in \autoref{fig:integrated_single_photons}. All recorded waveforms, triggered on the signal from the pulse generator, were integrated over a fixed time window, independent of whether a photon signal was present. As a result, the distribution includes events with zero detected photoelectrons, forming a pedestal peak centered around zero integrated signal, which corresponds to the electronic noise floor. The full distribution was modeled as a sum of Gaussian contributions corresponding to 0, 1, 2, and higher photoelectron events. The separation between adjacent peaks defines the single-photoelectron response, while the width of each Gaussian reflects the combined effects of electronic noise and gain fluctuations. While the width of the distribution for 0 p.e. is goverend by noise only, the width of the events with 1 or more p.e. also includes the photon response function of the \gls{pmt}. We verified this method to obtain the single photon light yield using an independent approach: by identifying single-photon pulses in individual traces across different LED intensities, we constructed a median pulse shape where the integral agrees well with the single-photoelectron response extracted from the multi-Gaussian fit.

The same measurements were repeated for the other two R9800-100 \glspl{pmt} that are used later in the timing experiments. While the single-photon light yield uncertainty of about \qty{32}{\percent} ($1\sigma$) drives the overall uncertainty of the light-yield calculation, the variation of the mean single-photon light yield among the three \glspl{pmt} was measured to be less than \qty{4}{\percent}.

\subsubsection{Absolute Light Yield Results}

The R9800-100 \gls{pmt} has a \gls{qe} of \qty{42}{\percent} at \qty{400}{\nano\meter}. At \qty{1300}{\volt}, the typical gain is about $1.1\times10^6$. The R9800-100 has been studied in detail by \citet{seng2013_R9800_study}. Additionally, Monte Carlo photon transport simulations were performed to account for photon-transport efficiency. \autoref{fig:LY_comparison_detected_pe} shows a direct comparison of the measured photoelectrons for each sample. Note that the \gls{zno} measurement was performed without a reflective mirror coating, whereas \gls{gan} and \gls{yap} employed reflective coatings. The integration window for \gls{gan} and \gls{zno} was \qty{25}{\nano\second} while \qty{165}{\nano\second} of the waveform were integrated for \gls{yap} due to the slower decay. The energy resolution of \qty{8.8}{\percent} \gls{fwhm} measured for \gls{zno} is noteworthy, given that the statistical limit is \qty{6.9}{\percent} for \gls{zno} corresponding to the statistical limit $2.355/\sqrt{N}$. Eq.~\eqref{eq:light_yield_mev_and_error} provides the formulas used to obtain the absolute light yields for alpha particles shown in \autoref{fig:LY_comparison_net}:
\begin{equation}
\label{eq:light_yield_mev_and_error}
\begin{aligned}
\mathrm{LY}_{\mathrm{MeV}} & = \frac{P}{\text{qe} \cdot T \cdot E},\\[2pt]
\Delta \mathrm{LY}_{\mathrm{MeV}} & = \mathrm{LY}_{\mathrm{MeV}}
\sqrt{\sum_{x\in\{P,\text{qe},T,E\}}\left(\frac{\Delta x}{x}\right)^2 }\,.
\end{aligned}
\end{equation}

Here, $P$ is the number of photoelectrons detected by the \gls{pmt}, $\text{qe}$ is the \acrlong{qe} of the \gls{pmt}, $T$ is the photon-transport efficiency from the scintillator to the \gls{pmt}, $E$ is the energy of the incident alpha particle in MeV, and $\mathrm{LY}_{\mathrm{MeV}}$ is the light yield in photons per MeV.

Although the single-photon response dominates the overall uncertainty, the measured value for \gls{yap} agrees well with the manufacturer’s reference. The monocrystalline \gls{zno} sample, as expected, significantly outperforms powder samples in light yield. The relative light yield is approximately 17\% for \gls{gan} and 43\% for \gls{zno} compared to \gls{yap}. One contributor to the lower detected-photon fraction for \gls{gan} is its high refractive index, which increases reflections at interfaces and narrows the total internal reflection acceptance cone. \autoref{tab:ly_comparison} summarizes the detected number of photons, material properties, and calculated light yield.

\begin{table}[htbp]
  \centering
  \small
  \setlength{\tabcolsep}{4pt} 
  \caption{Detected photoelectrons, refractive index, and calculated light yield for \gls{gan}, \gls{zno}, and \gls{yap}.}
  \label{tab:ly_comparison}
  \begin{tabular}{l
                  S[table-format=4.0]   
                  S[table-format=1.2]   
                  c}                    
    \toprule
    \textbf{Material} &
    \textbf{Detected p.e.} &
    \textbf{$n$} &
    \textbf{Calc.\ light yield (ph/MeV$_{\alpha}$)} \\
    \midrule
    GaN:Si &  390 & 2.51 & 1060 $\pm$ 350 \\
    ZnO:Ga & 1165 & 1.71 & 2600 $\pm$ 860 \\
    YAP:Ce & 4094 & 1.93 & 6090 $\pm$ 2010 \\ \bottomrule
  \end{tabular}
\end{table}
\FloatBarrier
\subsubsection{Ionoluminescence Spectra}
\begin{figure}
    \centering 
    \includegraphics[width=0.48\textwidth]{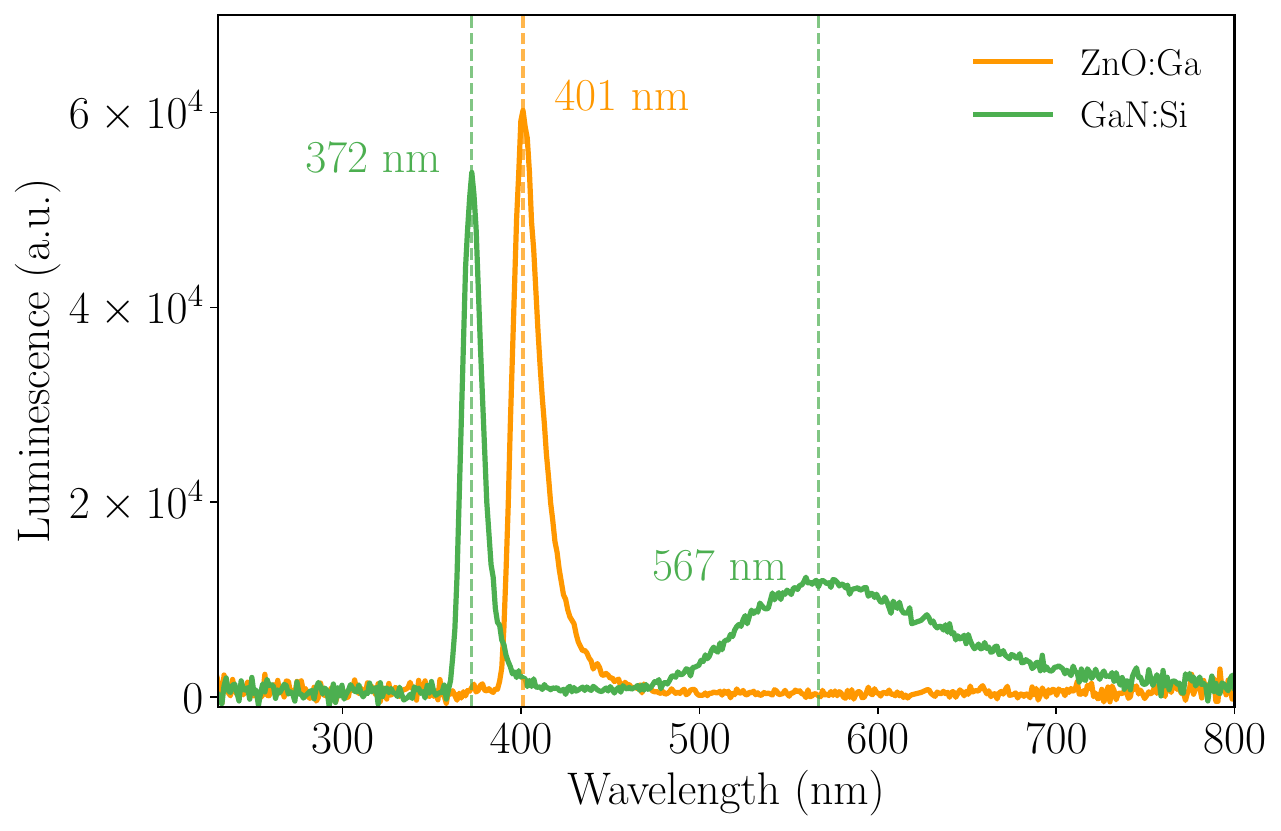}          
    \caption{\gls{am241} $\alpha$-induced ionoluminescence spectrum for \gls{gan} and \gls{zno}. 
    The emission peak of \gls{yap}, for reference, is at \qty{370}{\nano\meter}, according to the manufacturer Crytur. }\label{fig:IL_spectrum_gan_zno}
\end{figure}

Ionoluminescence is the emission of light from a material as a result of excitation by energetic ion irradiation, analogous to radioluminescence induced by X-rays or gamma rays.
Transmission ionoluminescence spectra were collected by irradiating the \gls{gan} and \gls{zno} samples with \SI{5.47}{\mega\electronvolt} $\alpha$-particles from a \SI{40}{\nano Ci} \gls{am241} source. The emitted light was analyzed using a SpectraPro-2150i spectrometer (Acton Research), coupled to a Pixis:1008 CCD camera (Princeton Instruments). In the transmission geometry, the ionoluminescence was detected from the side opposite to the irradiated face, ensuring that only light transmitted through the sample was measured. All spectra were corrected for the instrument response. The resulting ionoluminescence spectra for both materials are shown in \autoref{fig:IL_spectrum_gan_zno}.

While \gls{zno} exhibits a single strong emission around \qty{401}{\nano\meter}, \gls{gan} shows the direct bandgap emission at \qty{372}{\nano\meter} and, in addition, the greenish yellow luminescence peaking around \qty{567}{\nano\meter}, as also reported by \citet{yanagida2021photoluminescence_Gan} and \citet{TOMINAGA2026117697}. The contribution from the fast-bandgap emission is significantly higher than in the unintentionally doped GaN sample reported in \citet{yanagida2021photoluminescence_Gan}. The correlation between the fast decay and the \qty{372}{\nano\meter} emission was verified by \gls{tcspc} measurements with and without wavelength filters: with a \qtyrange{360}{380}{\nano\meter} filter, the \gls{gan} decay shows only the ultra-fast component, whereas the slow \qty{16}{\nano\second} component shown in \autoref{fig:GaN_TCSPC_trace} is absent. 
\FloatBarrier
\subsection{Coincidence Timing Resolution Results}

\subsubsection{Experimental Results}
\label{subsec:timing_results}
\begin{figure}
  \centering
  \begin{subfigure}{\linewidth}
      \centering
      \includegraphics[width=0.95\linewidth]{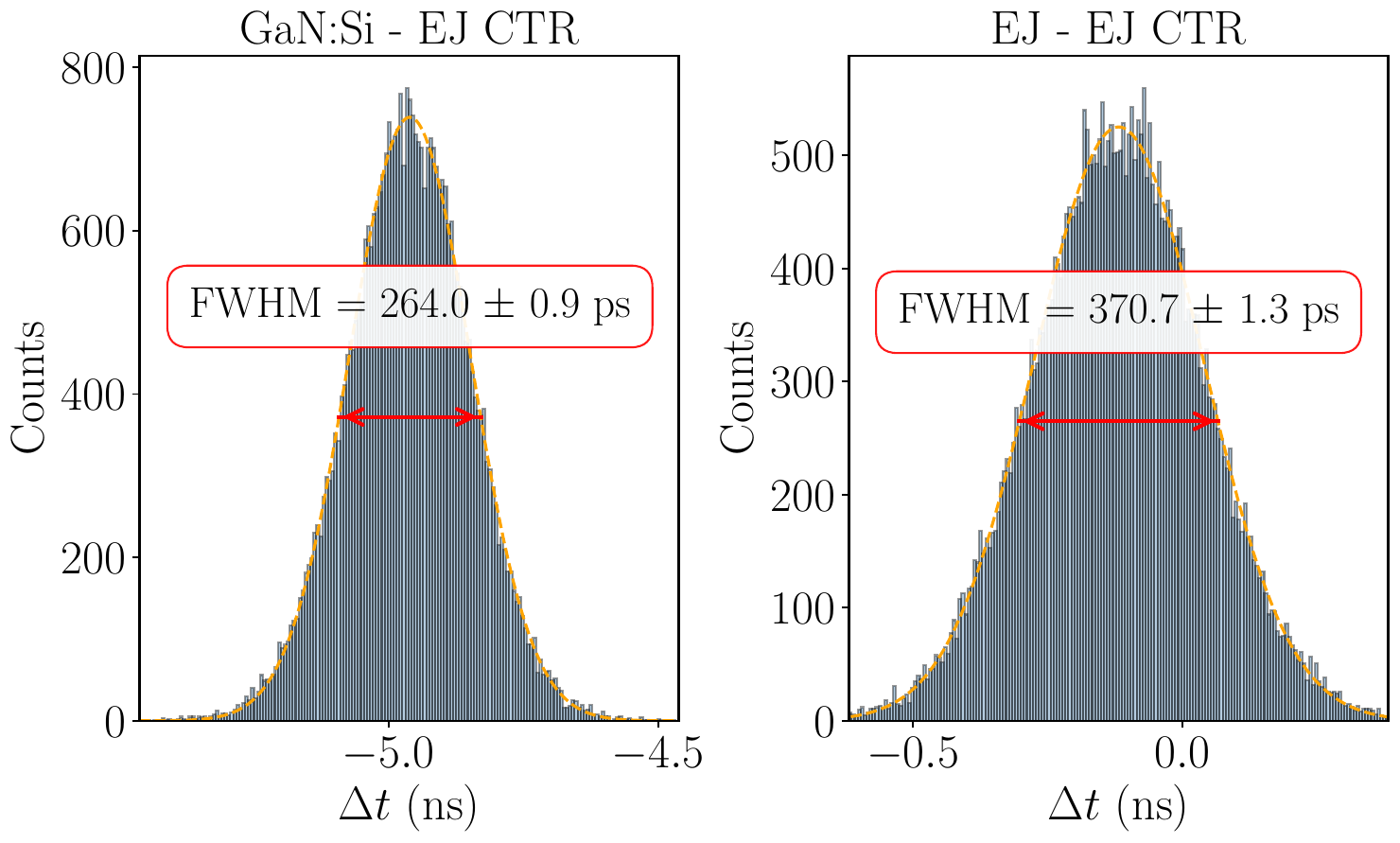}
      \caption{\gls{gan} \gls{ctr} distributions for coincident events.}
      \label{fig:ctr_gan}
  \end{subfigure}

  \vspace{1em}

  \begin{subfigure}{\linewidth}
      \centering
      \includegraphics[width=0.95\linewidth]{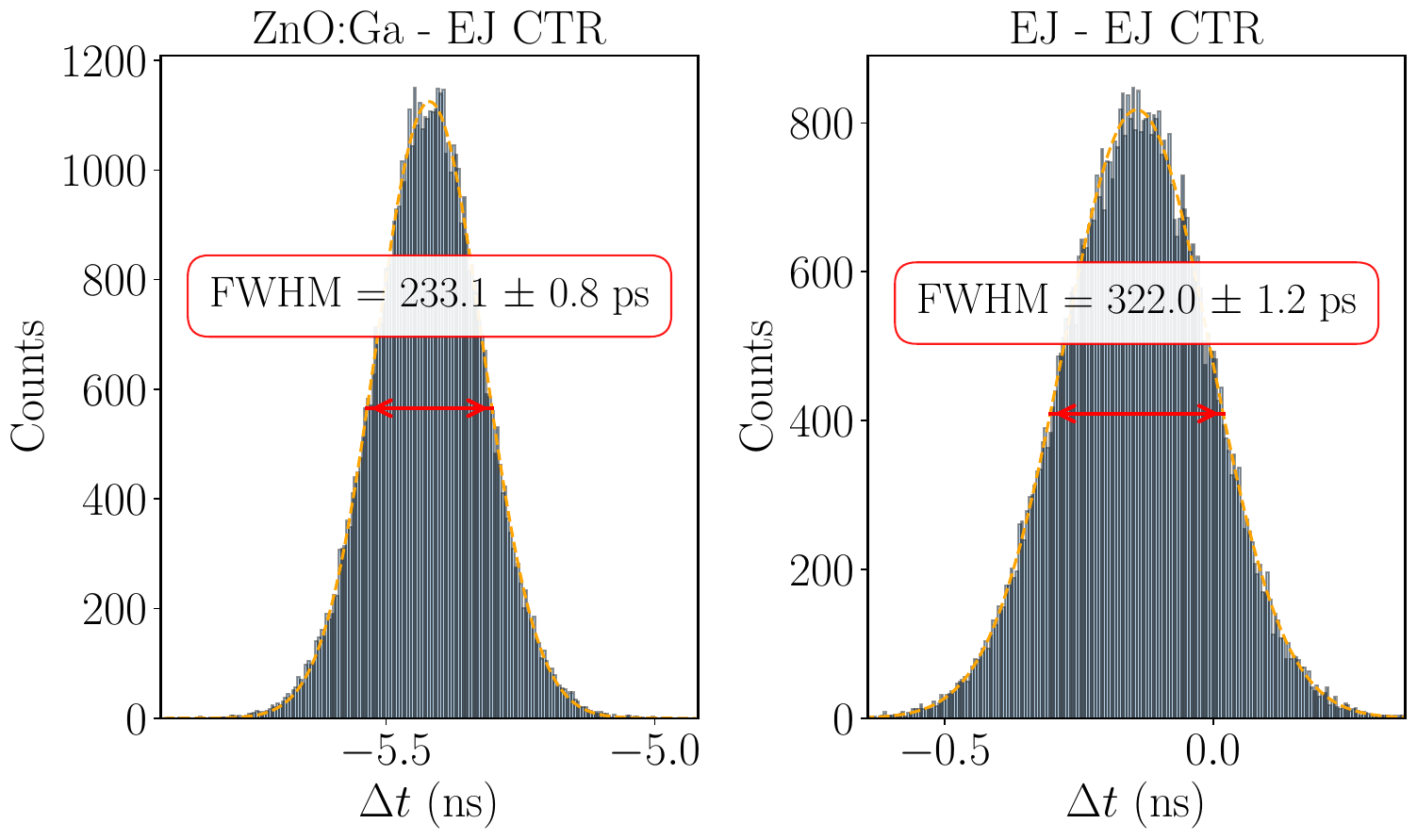}
      \caption{\gls{zno} \gls{ctr} distributions for coincident events.}
      \label{fig:ctr_zno}
  \end{subfigure}

  \vspace{1em}

  \begin{subfigure}{\linewidth}
      \centering
      \includegraphics[width=0.95\linewidth]{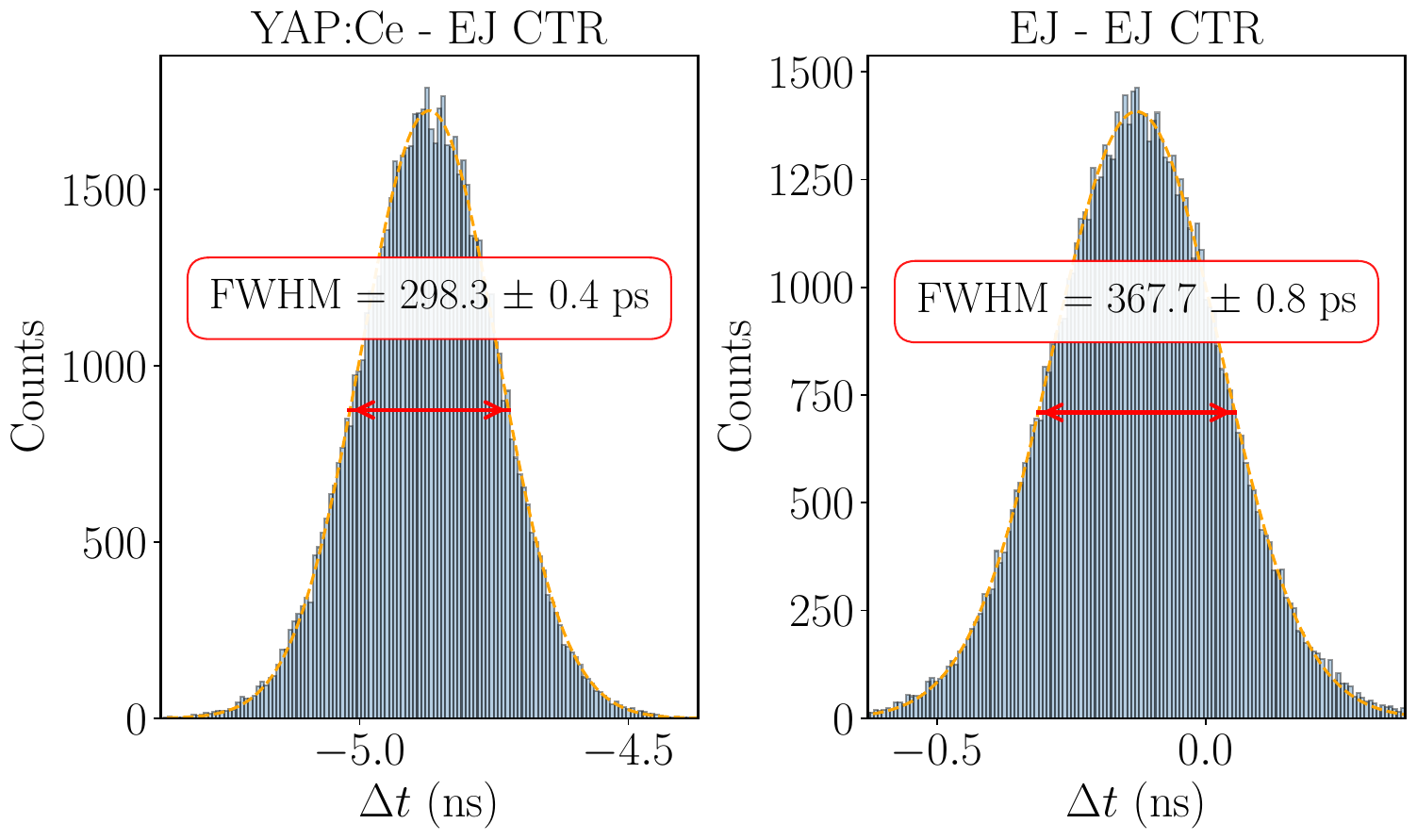}
      \caption{\gls{yap} \gls{ctr} distributions for coincident events.}
      \label{fig:ctr_yap}
  \end{subfigure}

  \caption{Coincidence timing resolution overview for all three scintillator materials. Shown is the timing uncertainty to detect each event in the sample scintillator and one EJ (left) and in both EJ-214 scintillators for reference (right).}
  \label{fig:ctr_all_materials}
\end{figure}

\begin{table}[htbp]
  \centering
  \small
  \setlength{\tabcolsep}{5pt}
  \caption{Photon selection windows and \gls{cfd} parameters used for timing measurements for each material and the reference EJ-214 scintillator.}
  \label{tab:timing_params}
  \begin{tabular}{c c c}
    \toprule
    \textbf{Material (PMT)} &
    \textbf{Photons [p.e.]} &
    \textbf{\gls{cfd} parameters} \\
    \midrule
    
      GaN:Si (1) & \qtyrange{210}{260}{} & d=\qty{1}{\nano\second}, f=\qty{0.53}{} \\
      EJ (2,3)   & \qtyrange{220}{260}{} & d=\qty{2.8}{\nano\second}, f=\qty{0.2}{} \\

    \midrule
      ZnO:Ga (1) & \qtyrange{700}{840}{} & d=\qty{0.8}{\nano\second}, f=\qty{0.32}{} \\
      EJ (2,3)   & \qtyrange{280}{330}{} & d=\qty{2.8}{\nano\second}, f=\qty{0.2}{} \\

    \midrule
      YAP:Ce (1) & \qtyrange{1900}{2650}{} & d=\qty{0.8}{\nano\second}, f=\qty{0.35}{} \\
      EJ (2,3)   & \qtyrange{230}{280}{}  & d=\qty{2.8}{\nano\second}, f=\qty{0.2}{} \\

    \bottomrule
  \end{tabular}
\end{table}

\begin{figure}
    \centering 
    \includegraphics[width=0.49\textwidth]{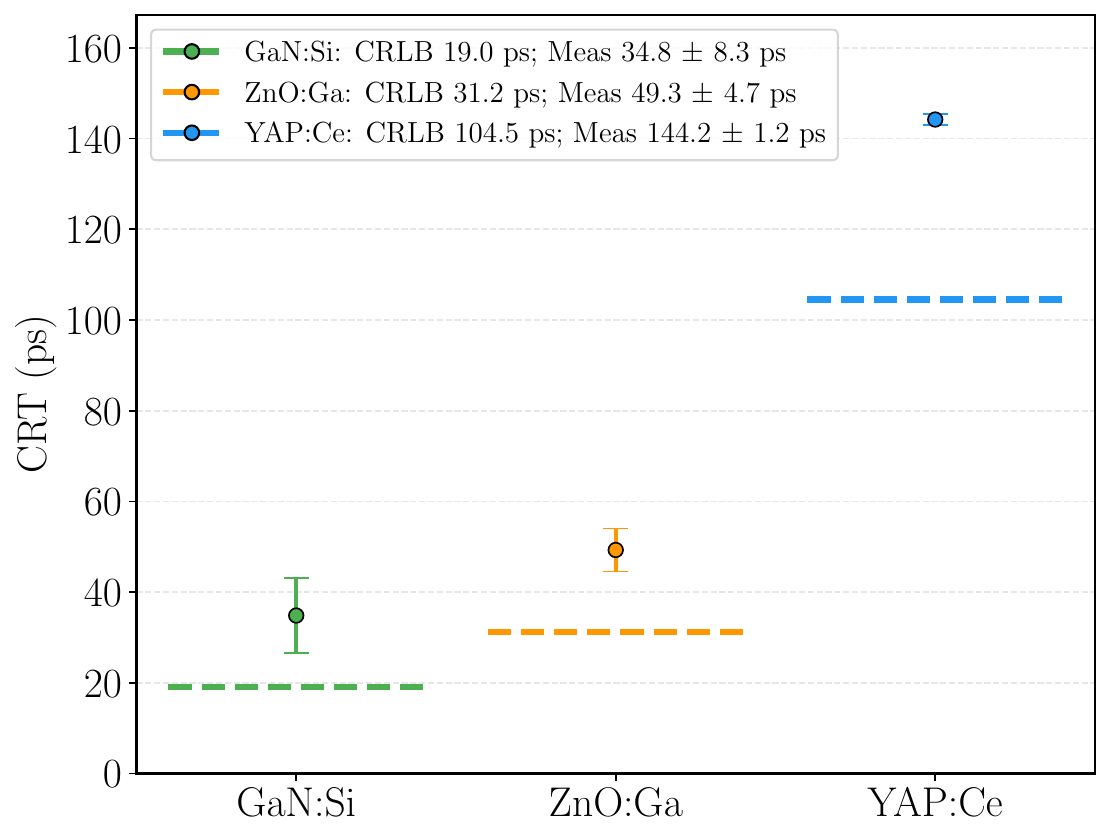}
    \caption{\gls{ctr} contribution comparison for \gls{gan}, \gls{zno}, and \gls{yap}. Also shown is the calculated \gls{crlb} for the Hamamatsu R9800-100 with the number of photons in the peak for each photon distribution in Figs.~\ref{fig:ctr_gan}, \ref{fig:ctr_zno}, and \ref{fig:ctr_yap}.}\label{fig:ctr_comparison}
\end{figure}

For all three samples, a total of \qty{200000}{} traces were collected in the triple coincidence setup described in \autoref{subsec:timing_setup}: one per channel for the two reference EJ-214-coupled \glspl{pmt} and the sample-coupled \gls{pmt}. Because the probability of accidental (random) coincidences is negligible with triple-coincidence triggering, no additional event selection was required. Figures~\ref{fig:ctr_gan},  \ref{fig:ctr_zno}, and \ref{fig:ctr_yap} show the timing uncertainty to detect individual alpha particle events in \gls{ctr} histograms for each sample (left) and the reference EJ-214 scintillators (right).To determine the trigger time in each channel, a custom \gls{cfd} was implemented based on the recorded trace data. The CFD was calculated using Eq. \eqref{eq:cfd}:
        \begin{equation}\label{eq:cfd}
            \text{CFD}(t) = S(t - d) - f \cdot S(t),
        \end{equation}
where $S(t)$ is one individual digitized and interpolated waveform, $d$ represents the delay and $f$ is the fraction. A \gls{cfd} optimization was performed to find the optimal parameters for each sample and the EJ-214 thin film. The optimized values are summarized in \autoref{tab:timing_params}. The table also captures the photon windows of the traces selected to measure the \gls{ctr}. As the \gls{crlb} and the experimentally achievable timing performance vary with the number of photons in each waveform, we selected narrow photoelectron windows (used as an energy proxy) for the signal in our collected traces. All three samples showed a Gaussian-like energy spectrum as seen in \autoref{fig:LY_comparison_detected_pe}, from which we selected an energy/photon window around the peak. However, in this setup, we now only have about \qty{1.9}{\mega\electronvolt} alpha particles instead of \qty{5.47}{\mega\electronvolt} alpha particles presented in \autoref{subsec:ly_results}, and therefore, the total number of detected photons is lower.

Time of flight of the alpha particles contributes only a constant offset between the mean photon-arrival times at the reference \glspl{pmt} and the sample \gls{pmt}. However, the event-to-event spread is governed by the convolution of the sample and EJ-214 \gls{ctr} contributions together with \gls{tts} and electronic jitter. To calculate the contribution of the sample scintillators, we use
\begin{equation}
\label{eq:ctr_contribution}
\begin{aligned}
\text{FWHM}_{\text{sample}} &= \sqrt{\text{FWHM}_{\text{sample+EJ}}^2 - \text{FWHM}_{\text{ref}}^2},\\
\text{where } \text{ FWHM}_{\text{ref}} &= \frac{\text{FWHM}_{\text{EJ+EJ}}}{\sqrt{2}},
\end{aligned}
\end{equation}
where FWHM$_{\text{EJ+EJ}}$ is the measured \gls{fwhm} of the \gls{ctr} distribution in the right panel of each figure, and FWHM$_{\text{sample+EJ}}$ is the measured \gls{fwhm} of the \gls{ctr} distribution on the left. The factor of $\sqrt{2}$ assumes that both EJ-coupled \glspl{pmt} contribute equally to the EJ+EJ timing distribution, which is justified by the measured variation of less than \qty{4}{\percent} in single-photon response among the three \glspl{pmt}. To compare FWHM$_{\text{sample}}$ with the theoretically possible \gls{crlb}, the photon-selection windows were chosen for each material to optimize the \gls{ctr} and its uncertainty. \autoref{fig:ctr_comparison} presents the resulting FWHM$_{\text{sample}}$ \gls{dtr} and the \gls{crlb} for each material. A central result is the measured \gls{dtr} of \qty{35 \pm 9}{\pico\second} for \gls{gan} and \qty{49 \pm 5}{\pico\second} for \gls{zno}. Note that for \gls{zno} this is similar to the result reported by \citet{cates2013zno_ga} for a powder \gls{zno} sample, although the \glspl{pmt} used in that work had a significantly faster \gls{tts}. This suggests that the higher photon yield of the single-crystal compensates for the slower \gls{tts} of the Hamamatsu R9800-100 \glspl{pmt}.

\subsubsection{Statistical Timing Resolution Limits}
\begin{figure}
  \centering

  \begin{subfigure}{\linewidth}
    \centering
    \includegraphics[width=\linewidth]{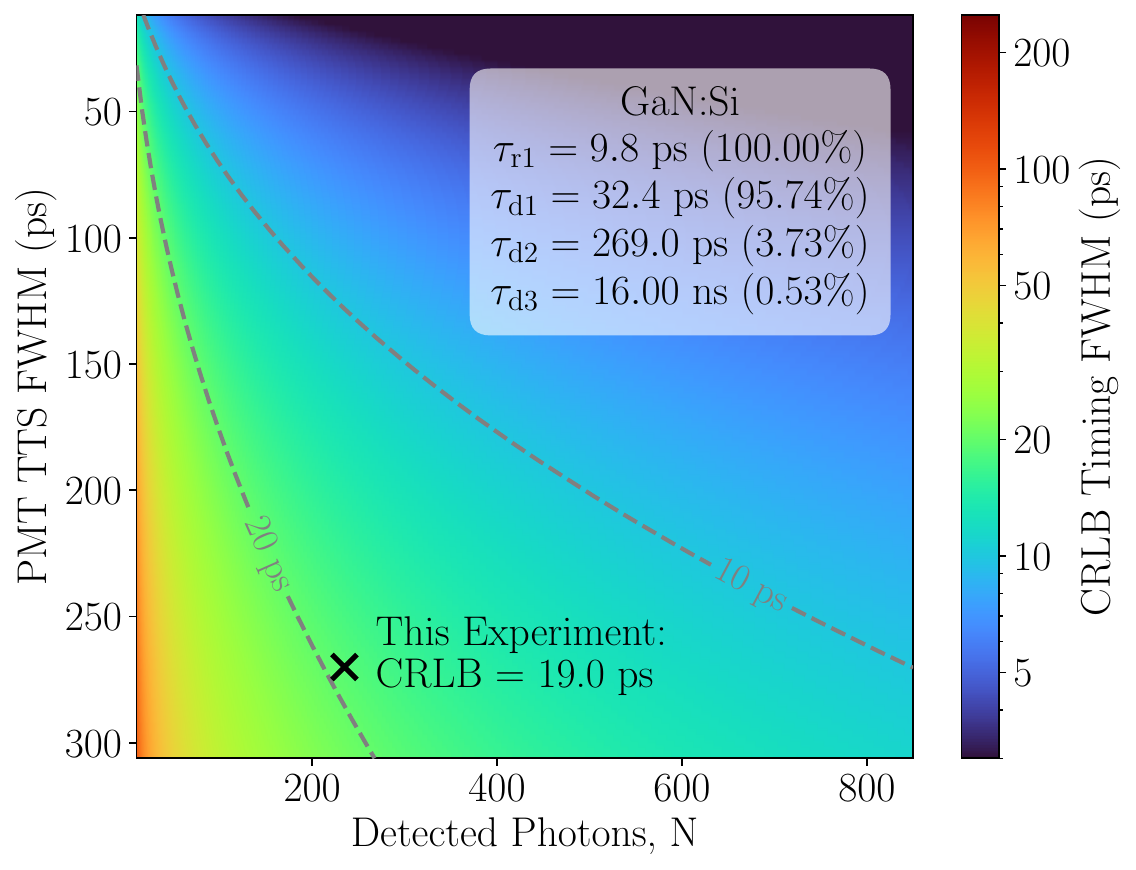}
    \caption{\gls{crlb} map for \gls{gan}: For \glspl{pmt} with sufficiently low \gls{tts}, \gls{dtr} of less then \qty{10}{\pico\second} are achievable.}
    \label{fig:crlb_sim_gan}
  \end{subfigure}
  \begin{subfigure}{\linewidth}
    \centering
    \includegraphics[width=\linewidth]{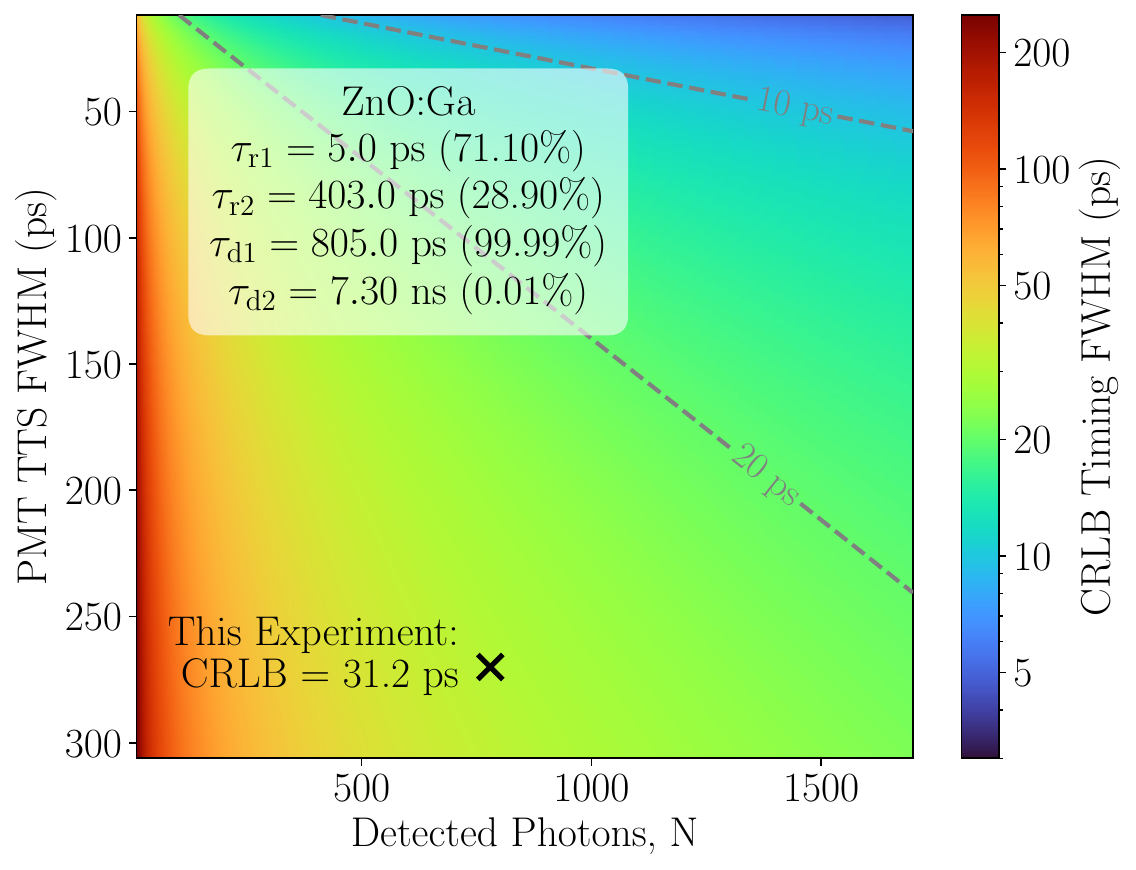}
    \caption{\gls{crlb} map for \gls{zno}: For \glspl{pmt} with sufficiently low \gls{tts}, \gls{dtr} down to \qty{20}{\pico\second} are achievable.}
    \label{fig:crlb_sim_zno}
  \end{subfigure}

  \caption{\gls{crlb} calculations based on the measured rise and decay constants of the \gls{gan} (top) and \gls{zno} (bottom) sample for a range of detected photons and different values for the \gls{pmt} \gls{tts}.}
  \label{fig:crlb_sim}
\end{figure}
To understand the prospects for even better timing performance with \gls{gan} and \gls{zno}, \autoref{fig:crlb_sim} shows the theoretically achievable \gls{crlb} for the \gls{dtr} as a function of detected photoelectrons and the \gls{tts} \gls{fwhm} of the \gls{pmt}. For this, the rise and decay times for both samples from \autoref{subsec:tcspc_results} were inserted into Eq.~\eqref{eq:cates1} to obtain the emission \gls{p.d.f.}, which was then used in Eq.~\eqref{eq:cates2} to calculate the \gls{crlb} via Eqs.~\eqref{eq:cates6}, \eqref{eq:cates7}, and \eqref{eq:fwhm_fisher}. With a photodetector with a \gls{tts} \gls{fwhm} below \qty{100}{\pico\second} and a similar \gls{qe} and therefore a comparable number of detected photons, timing resolutions on the order of \qtyrange{10}{20}{\pico\second} are theoretically achievable with \gls{gan} and \gls{zno}. Especially for \gls{gan}, one can obtain from \autoref{fig:crlb_sim_gan} the potential for a \gls{crlb} down to \qty{10}{\pico\second}, making it a very suitable candidate for ultra-fast timing applications. Note that the number of detected photons for the \gls{zno} sample could be significantly increased by adding a reflective layer as was done for the two other samples. 

\subsection{Position Resolution Results}
\label{subsec:posres_results}
\subsubsection{Experimental Results}
\begin{figure}
    \centering 
    \includegraphics[width=0.48\textwidth]{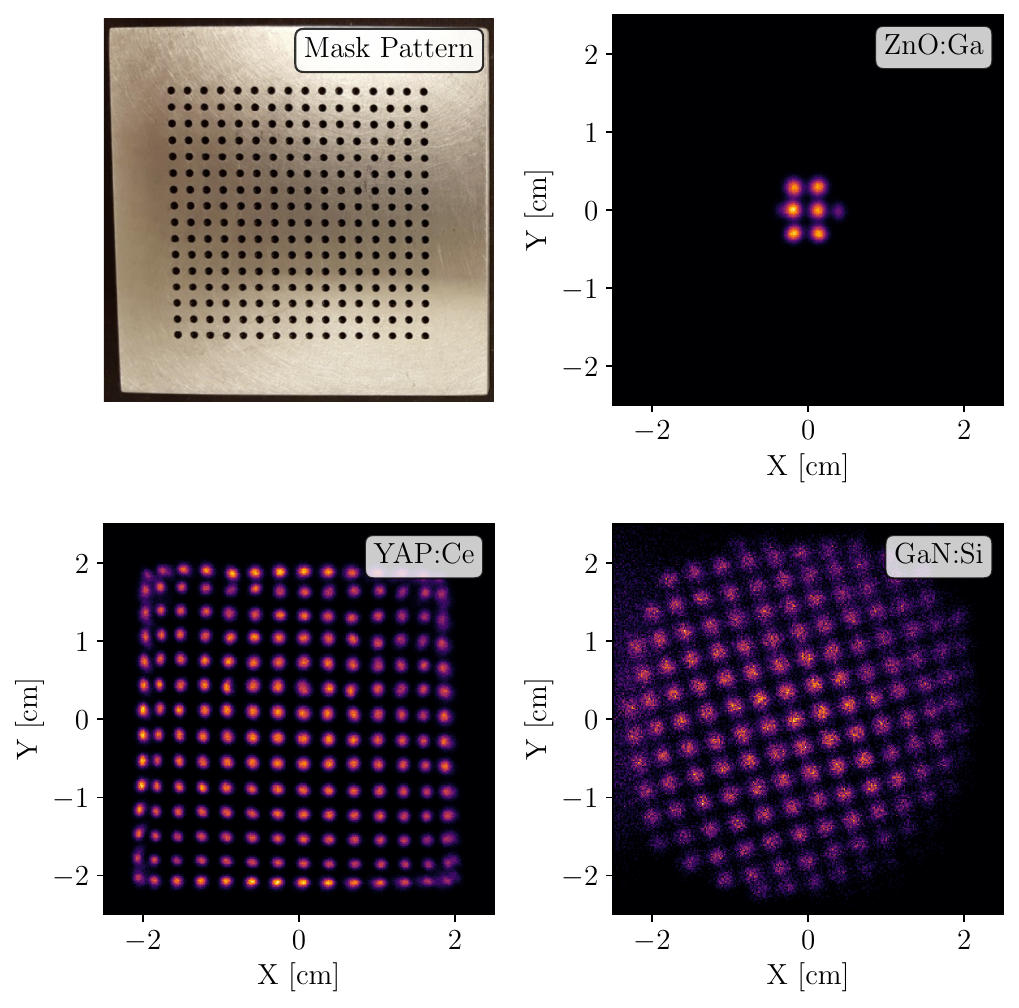}
    \caption{Qualitative position reconstruction measurement showing the ability to reconstruct a pattern (upper left) with all three samples. The different number of dots visible for each sample is due to the different sizes of each sample.}\label{fig:coarse_mask_measurements}
\end{figure}

\begin{figure}
  \centering
  \begin{subfigure}{\linewidth}
    \centering
    \includegraphics[width=\textwidth]{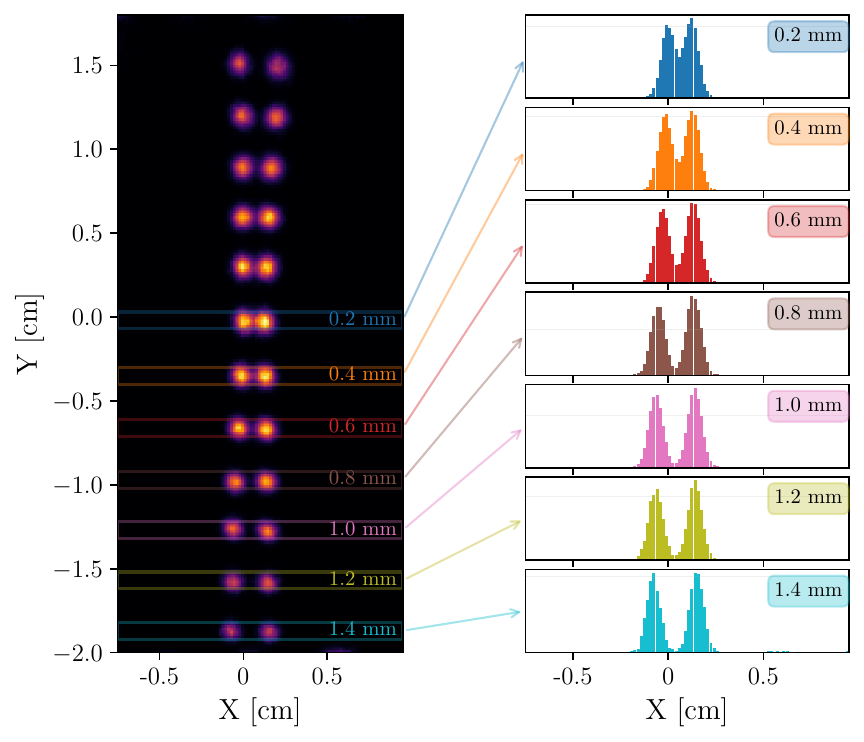}
    \caption{Reconstructed interaction positions for alpha particles in \gls{yap} (left) and a dimensionality-reduced histogram showing the intensity within the marked windows. A clear distinction is visible already at \qty{0.2}{\milli\meter}.}\label{fig:yap_fine_mask}
  \end{subfigure}
  \begin{subfigure}{\linewidth}
    \centering
    \includegraphics[width=\textwidth]{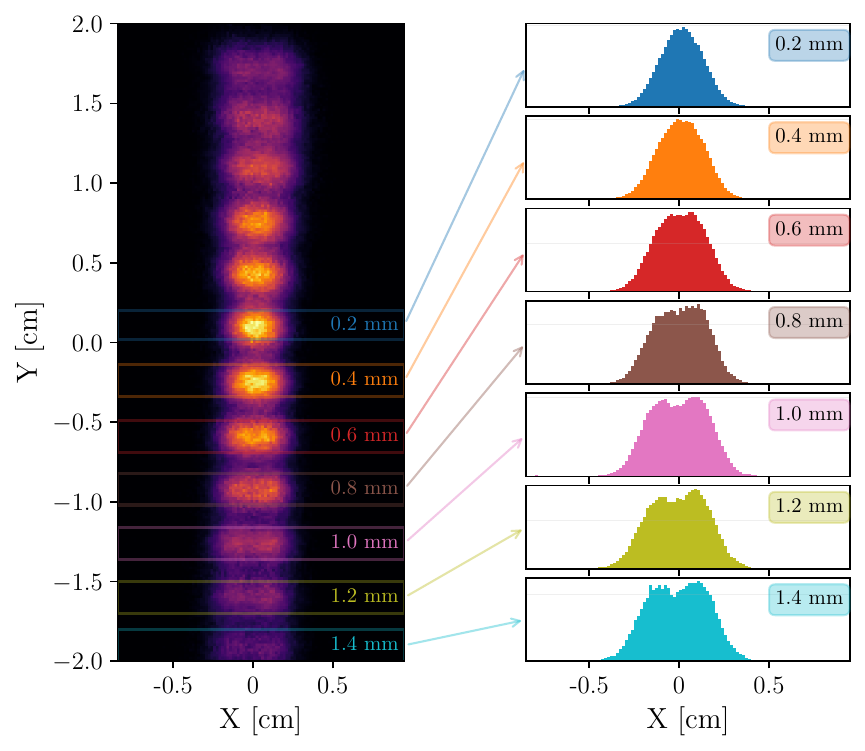}
    \caption{Reconstructed interaction positions in \gls{gan} (left) and histogram showing the intensity within the marked windows. A slight distinction emerges at \qty{0.8}{\milli\meter}. At \qty{1}{\milli\meter}, both holes are sufficiently resolved.}\label{fig:gan_fine_mask}
  \end{subfigure}
  \caption{Quantitative position-resolution measurements using a mask with hole-to-hole distances between \qty{0.2}{\milli\meter} and \qty{1.4}{\milli\meter}. Measurements for \gls{yap} (a) and \gls{gan} (b).}
  \label{fig:fine_mask_resolution}
\end{figure}
\autoref{fig:coarse_mask_measurements} shows a qualitative comparison of the position-resolution capabilities for \gls{zno}, \gls{gan}, and \gls{yap}. Because total internal reflection at the sample–vacuum interface is a major loss mechanism, the light yield in this geometry is significantly lower than in \autoref{subsec:ly_setup}. We measured approximately 25\% of the photoelectron yield obtained in the light yield setup \autoref{subsec:ly_results}: about \qty{1000}{} p.e. for YAP and \qty{90}{} p.e. for GaN with the H13700-03. A four-channel PicoScope D6404 with \qty{500}{\mega\hertz} bandwidth and a sampling rate of \qty{1.25}{GS/s} was used to acquire the data. For \gls{zno} (upper right), only a small fraction of the H13700-03 \gls{pmt} was covered with the scintillator, and therefore only seven grid points are visible. For \gls{yap} (lower left), surrounding grid points are distorted due to edge effects in the \gls{pmt} (see e.g. \citet{ayllon2021_api_3d_recon} for more details on edge effects). \gls{gan} (lower right) shows a coarser image with reduced resolution. The sample's round shape is clearly visible.

Additionally, a finer mask was placed inside the vacuum chamber on top of the \gls{gan} and \gls{yap} samples to replace the coarse mask for a more quantitative assessment of the position resolution. The \gls{zno} sample could not be measured with this mask because its small lateral size (\qtyproduct{10 x 10}{\milli\meter}) did not extend to the fine mask holes, which are located off-center in the field of view. The measurements with that mask, which has hole-to-hole distances from \qty{0.2}{\milli\meter} to \qty{1.4}{\milli\meter}, are shown in \autoref{fig:fine_mask_resolution}. In agreement with \citet{ayllon2021_api_3d_recon}, the position resolution for \gls{yap} is at least \qty{0.2}{\milli\meter}, whereas resolving two holes with \gls{gan} is noticeably more difficult. We note that \gls{gan} achieves a position resolution of approximately \qty{1}{\milli\meter}, which is adequate for many \gls{api} imaging scenarios.

\FloatBarrier
\subsubsection{Position Resolution Simulations}
\label{subsec:pos_res_sim}
\begin{figure}
    \centering 
    \includegraphics[width=0.4\textwidth]{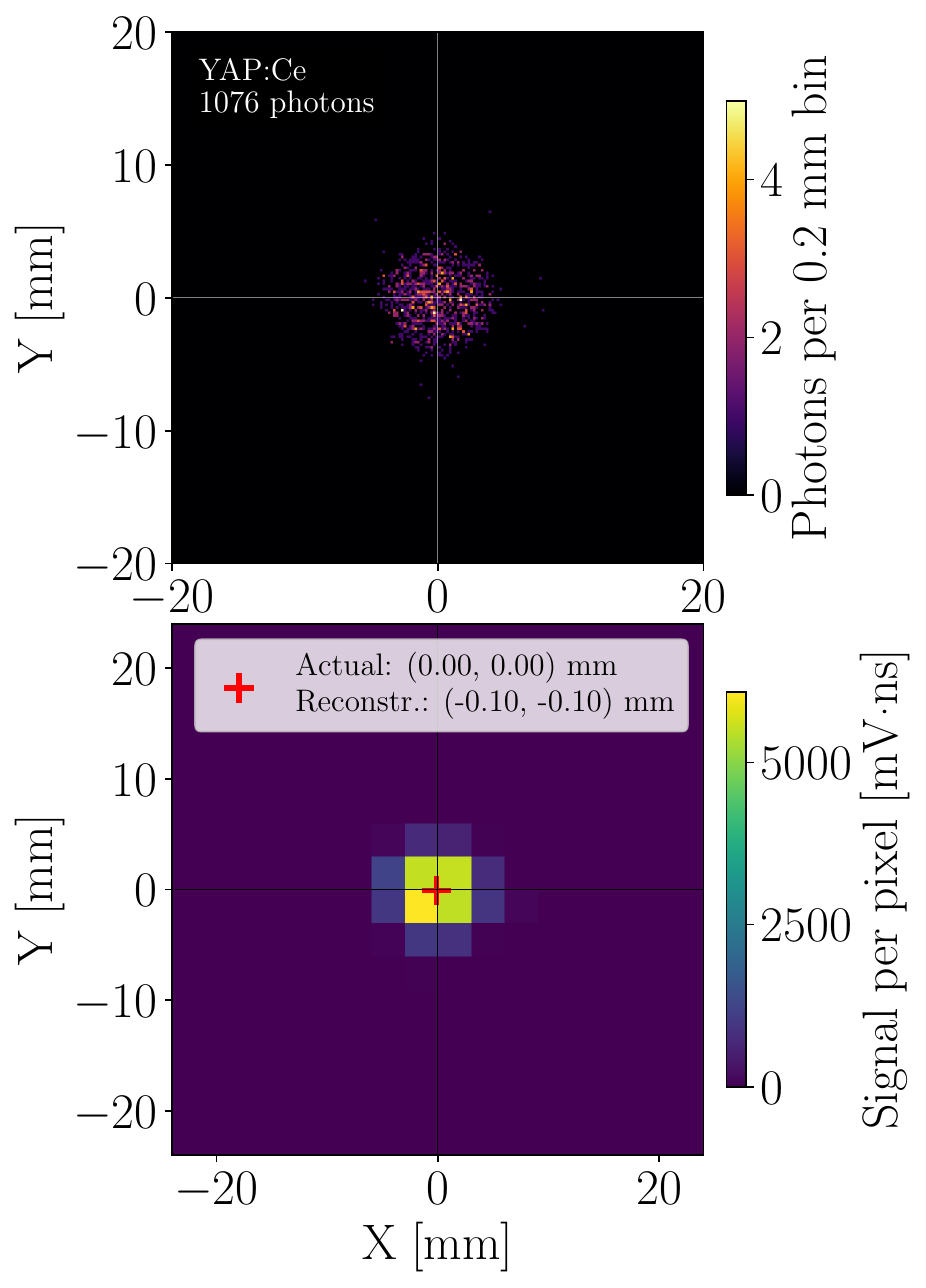}
    \caption{One simulated \qty{3.5}{\mega\electronvolt} alpha-particle interaction in \gls{yap}: simulated photon-arrival distribution at the \gls{pmt} (left), Hamamatsu H13700 \gls{pmt} response per pixel, and resulting reconstruction of the interaction location (right).}\label{fig:yap_alpha_single_event_sim}
\end{figure}

\begin{figure}
    \centering 
    \includegraphics[width=0.48\textwidth]{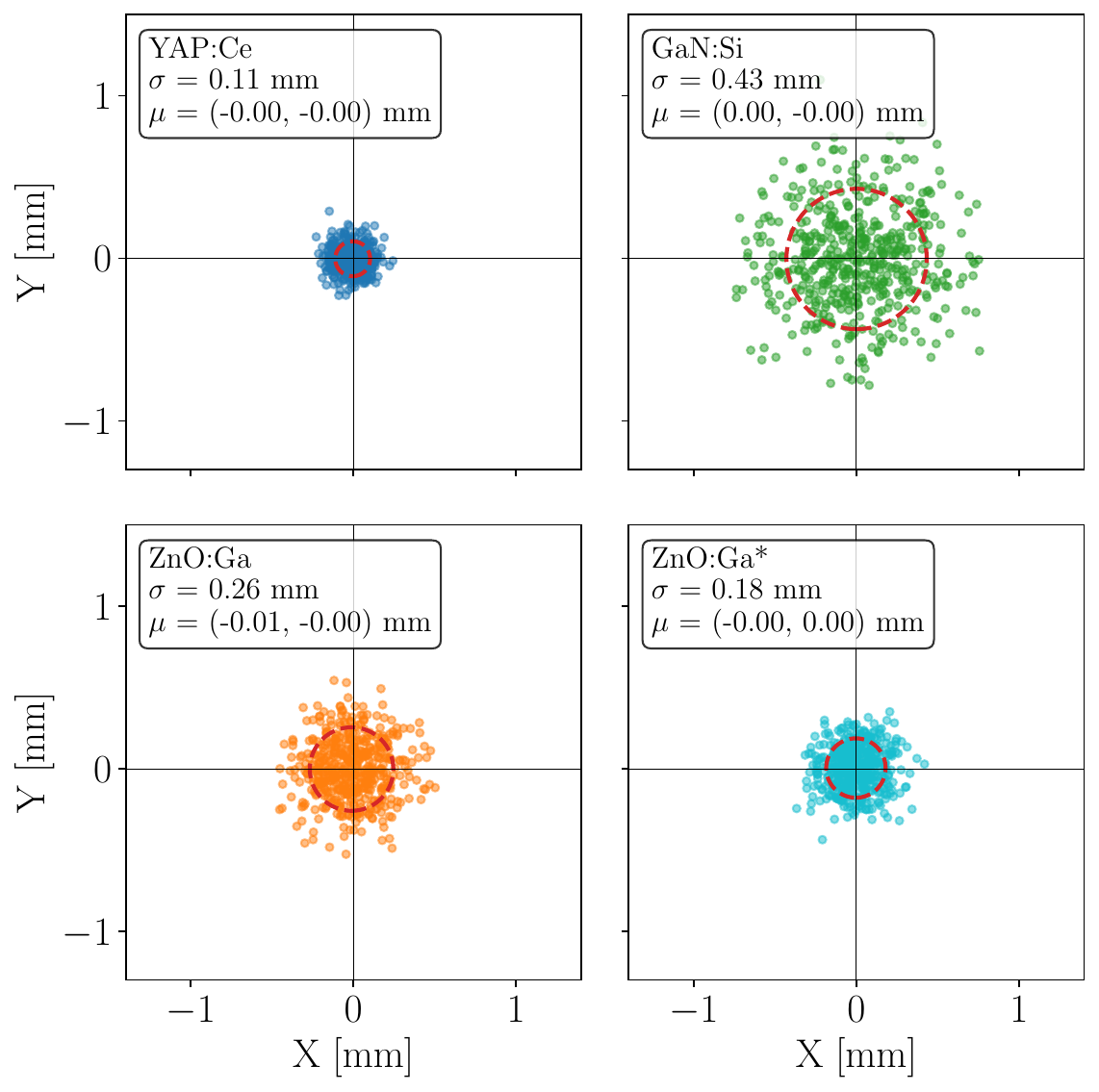}
    \caption{Simulation of 500 individual \qty{3.5}{\mega\electronvolt} alpha-particle interactions in \gls{yap}, \gls{gan}, \gls{zno}, and a thinner \qty{100}{\micro\meter} \gls{zno}* scintillator with a reflective aluminum coating for better comparison with \gls{yap} and \gls{gan}.}\label{fig:pos_resulution_simulation}
\end{figure}

To quantify the position-resolution capabilities of the \gls{zno} sample, we performed Monte Carlo simulations using the same Python code as in \autoref{subsec:ly_setup}. For each sample, the number of photons was drawn from the measured distribution. In \autoref{fig:yap_alpha_single_event_sim}, the upper panel shows the distribution of photons arriving at the \gls{pmt} after the photon-transport simulation. In contrast, the lower panel shows the integrated per-pixel signal and the reconstruction derived from it. The H13700-03 has a pixel size of \qtyproduct{3 x 3}{\milli\meter}. The simulation was repeated 500 times for each sample. \autoref{fig:pos_resulution_simulation} shows the distribution of reconstructed positions, which ideally cluster at the origin. The spread of this distribution serves as a measure of the achievable position resolution.

For \gls{yap} and \gls{gan}, the results agree with the measurements in \autoref{subsection:posresolution_exp_setup}. The slightly worse measured position resolution for \gls{gan} can be explained by the data acquisition system, which reduced energy resolution and therefore spatial resolution due to the comparatively low sampling rate of \qty{1.25}{GS/s}, where the pulse shape of a \gls{gan} signal is sampled by only a few points. For \gls{zno}, two simulations were run: the first represents the sample used in this work and shows that, like the measured light yield, the achievable position resolution is between \gls{yap} and \gls{gan}, closer to the performance of \gls{gan}. However, when a thinner sample (\qty{100}{\micro\meter}, reducing self-absorption) was used and coated with a mirror layer on one side, the light yield increased substantially, improving the position resolution by more than 30\% to about \qty{0.2}{\milli\meter}.

\section{Conclusions}
We characterized single-crystal \gls{zno} and near-single-crystal \gls{gan} for in-vacuum alpha-particle detection, demonstrating fast timing, energy-resolved alpha peaks, and high position resolution relevant to \gls{api} applications. To our knowledge, the use of \gls{gan} in \gls{api} systems is novel. We show that these crystals achieve a $\ge$$5\times$ improvement in timing resolution over \gls{yap}, a commonly used high-performance scintillator in such systems. Using \gls{tcspc}, we extracted rise and decay components. We quantified absolute and relative light yields, and measured the \gls{ctr} as well as the resulting \gls{dtr} of our samples in an alpha particle triple-coincidence setup. The main experimental results are summarized in \autoref{tab:scintillator_summary}: (i) primary decay components of \qty{32.4\pm1}{\pico\second} for \gls{gan}, \qty{805\pm1}{\pico\second} for \gls{zno} and \qty{24.3\pm0.1}{\nano\second} for \gls{yap}, (ii) absolute light yield of \qty{1060 \pm 350}{ph/\mega\electronvolt}$_\alpha$ (\gls{gan}), \qty{2600 \pm 860}{ph/\mega\electronvolt}$_\alpha$ (\gls{zno}), \qty{6090 \pm 2010}{ph/\mega\electronvolt}$_\alpha$ (\gls{yap}), (iii) \gls{dtr} of \qty{35 \pm 9}{\pico\second} for \gls{gan} and \qty{49 \pm 5}{\pico\second} for \gls{zno} compared to \qty{144\pm2}{\pico\second} for \gls{yap}, and (iv) position resolution of at least \qty{0.2}{\milli\meter} for \gls{yap}, about \qty{1}{\milli\meter} for \gls{gan}, and expected \qty{0.3}{\milli\meter} for \gls{zno}, and down to \qty{0.18}{\milli\meter} with a reflective layer in a thinner geometry. Our \gls{crlb} analysis indicates that with \glspl{pmt}, \glspl{mcp} or silicon-photomultiplier achieving \gls{tts} \gls{fwhm} $<\!\qty{100}{\pico\second}$ and comparable \gls{qe}, \gls{gan} and \gls{zno} could, in principle, support \gls{dtr} near or even below \qty{10}{\pico\second}.

\begin{table}
\centering
\small
\setlength{\tabcolsep}{4pt} 
\caption{Summary of experimental results for \gls{gan}, \gls{zno}, and \gls{yap} scintillators. We present their primary decay components, light yields, contributions to coincidence timing resolution, and position reconstruction capabilities.}
\label{tab:scintillator_summary}
\begin{tabular}{lccc}
\toprule
 & \textbf{GaN:Si} & \textbf{ZnO:Ga} & \textbf{YAP:Ce} \\
\midrule
Primary $\tau_d$ [\si{\nano\second}] & $0.032 \pm 0.001$ & $0.805 \pm 0.001$ & $24.3 \pm 0.1$ \\
LY [ph/\si{\mega\electronvolt}$_\alpha$] & $1060 \pm 350$ & $2600 \pm 860$ & $6090 \pm 2010$ \\
$\lambda_{\mathrm{em}}$ peak [\si{\nano\meter}] & 372 & 401 & 365 (Crytur) \\
CTR [\si{\pico\second}] & $35 \pm 9$ & $49 \pm 5$ & $144 \pm 2$ \\
Pos. res. [\si{\milli\meter}] & $\sim 1$ & $\sim 0.3$ (est.) & $\le 0.2$ \\
\bottomrule
\end{tabular}
\end{table}

From a systems perspective, faster decay and reduced pile-up open a path to higher operational rates and finer depth resolution in an integrated \gls{api} system. While accidental coincidences remain a limiting factor at elevated neutron rates, improved \gls{ctr} can narrow the coincidence window for certain applications where it is limited by the \gls{ctr} and not neutron time of flight within the region of interest, and can therefore lower the random-coincidence rate. 

Practical limitations are self-absorption and total internal reflection at sample-vacuum interfaces, which reduce the effective light yield (especially for \gls{zno}/\gls{gan} near the band edge). At the same time, optical design (thinning, reflective layers, index matching) measurably improves both collection efficiency and position resolution. 

\addtolength{\textheight}{-1cm}   





\section*{Data availability}
All data used in this work and all scripts to recreate the plots shown in this work can be found on Zenodo \cite{zenodo}.

\section*{Acknowledgement}

We thank Michal Boćkowski from the Institute of High Pressure Physics of the Polish Academy of Sciences for providing the \gls{gan} sample. We also want to thank Takeshi Katayanagi for technical support at LBNL.\\
This work was supported by the NASA PICASSO program under grant No. 22-PICASSO22-0013. Work at Lawrence Berkeley National Laboratory was conducted under U.S. Department of Energy Contract No. DE-AC02-05CH11231.


\printcredits

\bibliographystyle{unsrtnat}

\bibliography{cas-refs}


\end{document}